\documentclass[aps,prd,twocolumn,showpacs,floatfix,preprintnumbers,amsmath,amssymb,nofootinbib]{revtex4}
\input epsf
\usepackage{graphicx,color}

\def\VEV#1{\left\langle #1 \right\rangle}

\newcommand{\apjl}{Astrophys. J. Lett.}

\newcommand{\apjs}{Astrophys. J. Suppl. Ser.}

\newcommand{\mnras}{Mon. Not. R. Astron. Soc.}

\newcommand{\jcap}{JCAP.}
\newcommand{\jhep}{J. High Energy Phys.}

\newcommand{\prep}{Phys. Rep.}
\newcommand{\jtep}{Sov. Phys. JETP.}
\newcommand{\plb}{Phys. Lett. B.}

\def\rmX{\mathrm{X}}
\def\rmT{\mathrm{T}}
\def\rmB{\mathrm{B}}
\def\rmE{\mathrm{E}}

\def\VEV#1{\left\langle #1 \right\rangle}

\newcommand{\lsub}[2]{ \ensuremath{{}_{#2}\!{#1}}}

\newcommand{\lsim}{\mathrel{\hbox{\rlap{\lower.55ex\hbox{$\sim$}} \kern-.3em \raise.4ex \hbox{$<$}}}}
\newcommand{\gsim}{\mathrel{\hbox{\rlap{\lower.55ex\hbox{$\sim$}} \kern-.3em \raise.4ex \hbox{$>$}}}}
\def\wigner#1#2#3#4#5#6{ \left( \begin{array}{ccc} #1 & #3 & #5
\\ #2 & #4 & #6 \\ \end{array} \right)}

\def\tenspher#1#2#3#4{ \left( \begin{array}{cc} #1 & #3 
\\ #2 & #4  \\ \end{array} \right)}
\begin{document}
\title{Compensated Isocurvature Perturbations and the Cosmic
     Microwave Background}
\author{Daniel Grin$^1$, Olivier Dor\'e$^{2,3}$, and Marc Kamionkowski$^2$}
\affiliation{$^1$School of Natural Sciences, Institute for Advanced
     Study, Princeton, New Jersey 08540}
\affiliation{$^2$California Institute of Technology, Mail Code 350-17,
     Pasadena, California 91125}
\affiliation{$^3$Jet Propulsion Laboratory, California Institute
     of Technology, Pasadena, California 91109} 
\date{\today}
\begin{abstract}
Measurements of cosmic microwave background (CMB) anisotropies
constrain isocurvature fluctuations between photons and
nonrelativistic particles to be subdominant to adiabatic
fluctuations. Perturbations in the relative number densities of
baryons and dark matter, however, are
surprisingly poorly constrained. In fact, baryon-density
perturbations of fairly large amplitude may exist if they are
compensated by dark-matter perturbations, so that the total
density remains unchanged. These compensated isocurvature
perturbations (CIPs) leave no imprint on the CMB at observable
scales, at linear order. B modes in the CMB
polarization are generated at reionization through the modulation of the optical depth by CIPs, but this induced
polarization is small.  The strongest known constraint $\lesssim 10\%$
to the CIP amplitude comes from galaxy-cluster baryon fractions.
Here it is shown that modulation of the baryon density by 
CIPs at and before the decoupling of Thomson scattering at $z\sim 1100$ gives rise to CMB effects
several orders of magnitude larger than those considered before.  Polarization B modes are induced, as are correlations between temperature/polarization
spherical-harmonic coefficients of different $lm$.
It is shown that the CIP field at the surface of last scatter can be
measured with these off-diagonal correlations.  The
sensitivity of ongoing and future experiments to these
fluctuations is estimated. Data from the WMAP, ACT, SPT, and Spider experiments will be sensitive to fluctuations with amplitude $\sim5-10\%$. The Planck satellite and Polarbear experiment will be sensitive to fluctuations with amplitude $\sim3\%$.  SPTPol, ACTPol, and future space-based polarization methods will probe amplitudes as low as $0.4\%-0.6\%$. In the cosmic-variance limit, the smallest CIPs that could be detected with the CMB are of amplitude $\sim0.05\%$.

\end{abstract}
\maketitle

\section{Introduction}\label{iref}

The concordance cosmological model posits a nearly
scale-invariant spectrum of primordial density fluctuations with
adiabatic initial conditions, for which the ratios of
neutrino, photon, baryon, and dark-matter number densities are
homogeneous.  The simplest inflationary models predict adiabatic
fluctuations
\cite{Guth:1982ec,Linde:1982uu,bardeen_iso,Hawking:1982cz,mukhanov_iso,starobinsky_iso},
and adiabatic
fluctuations are consistent with measurements of cosmic
microwave background (CMB) temperature/polarization anisotropies
\cite{wmap7_komatsu_a} and the clustering of galaxies
\cite{sdss_lrg,sdss_pspec}.

Isocurvature perturbations are fluctuations in the ratios of
number densities of various particle species.  They are produced
in topological-defect models for structure formation
\cite{branden_topo} and in more complicated models of inflation
\cite{linde_iso,linde_mukhanov,mukhanov_b,axenides,langlois_riazuelo,langlois_double_iso}.
CMB temperature anisotropies limit the amplitude of baryon
isocurvature perturbations (fluctuations in the
baryon-to-photon ratio)
\cite{peebles_iso_a,peebles_iso_b} and CDM
isocurvature perturbations (fluctuations in the
dark-matter--to--photon ratio) \cite{burns_axion,seckel_turner,hu_iso_a,hu_iso_b} to be 
$\lesssim13\%$ of the total perturbation amplitude
\cite{boomerang_iso_limits,planck_iso,wmap7_komatsu_a,wmap7_komatsu_b,tommaso,zunckel_iso,bucher_dunkley_a,kawasaki_iso,beltran_viel_iso,seljak_iso,riazuelo_iso_constraint,takahashi_iso_nongauss}.

Our intuition thus suggests the matter in the early
Universe was very smoothly distributed.  It therefore comes as
somewhat of a surprise to learn that perturbations in the baryon
density can be almost arbitrarily large, as long they are
compensated by dark-matter
perturbations such that the total nonrelativistic matter density
remains unchanged \cite{gordon_pritchard,holder}. These compensated isocurvature perturbations (CIPs) thus obey \begin{eqnarray}
\rho_{\rm c}\delta_{\rm c}^{\rm CI}+\rho_{\rm b}\delta^{\rm CI}_{\rm b}=0,~~~\delta_{\gamma}^{\rm CI}=0,\end{eqnarray} where $\delta_{\rm c}$, $\delta_{\rm b}$, and $\delta_{\rm \gamma}$ are fractional energy density perturbations in the dark matter, baryons, and photons, respectively, while $\rho_{\rm c}$ and $\rho_{\rm b}$ are the homogeneous dark matter and baryon densities.

CIPs induce no curvature perturbation at early times, and they therefore leave the photon density---and thus large-angle CMB
fluctuations---unchanged at linear order.  CIPs induce baryon motion through baryon-pressure gradients, but these motions occur only at the baryon sound speed which, at the time when Thomson scattering first decouples ($z\sim 1100$, \textit{decoupling} hereafter), is $(v/c)\sim (T/m_p)^{1/2}\sim
(\mathrm{eV}/\mathrm{GeV})^{1/2}\sim 10^{-4.5}$.  The effects of
these motions on CMB temperature and polarization anisotropies
thus occur only on distances smaller than $\sim10^{-4.5}$ times
the sound horizon at decoupling or CMB multipole moments
$l\sim 10^6$
\cite{gordon_pritchard,gordon_lewis_curvaton,challinor_compensate},
scales far smaller than those probed by CMB experiments.

The effect of CIPs on galaxy surveys is also believed to be small
\cite{gordon_pritchard}.  Big-bang nucleosynthesis (BBN) and
galaxy-cluster baryon fractions constrain the CIP
perturbation amplitude to be $\lesssim10\%$ \cite{holder}.
Measurements of fluctuations in 21-cm radiation from
atomic hydrogen during the dark ages may be sensitive to these
perturbations
\cite{barkana,challinor_compensate,gordon_pritchard,sekigu}, but
these measurements are a long way in the future. 

In Ref.~\cite{holder}, it was shown that, although CIPs produce
no observable effect on the CMB at linear order in perturbation
theory, they modulate the CMB fluctuations produced by
adiabatic perturbations.  In particular, it was shown that B
modes in the CMB polarization are produced by the angular modulation
in the reionization optical depth induced by the CIP.

Here, we consider the additional effects on CMB fluctuations that
arise from modulation of the baryon density by CIPs at and before
decoupling.  CIPs modulate the free-electron density. They thus change the photon diffusion length and thickness of the surface of last scattering (SLS) on different patches of sky. CIPs also change the weight of the baryon-photon plasma and thus the details of the acoustic-peak structure in the CMB power spectrum.  Variation in the baryon density from one region on the
sky to another thus leads to a modulation of the small-scale
power spectrum from one region of sky to another. This
induces B modes in the polarization and nontrivial
higher-order correlations in the temperature/polarization map
analogous to those induced by variations of other
cosmological parameters \cite{sigurdson_alpha} and those induced
by weak gravitational lensing \cite{lewis_lens_review}.

As we show below, the effects of CIPs on CMB fluctuations from
decoupling are several orders of magnitude larger than those from reionization,
and so the CMB should provide a far more sensitive probe of CIPs
than envisioned in Ref.~\cite{holder}.  We therefore follow
through and develop the formalism required to look for CIPs with
the CMB.  To do so, we write down the minimum-variance
estimators that can be constructed from a CMB
temperature-polarization map for the CIP field $\Delta(\hat n)$
as a function of position $\hat n$ on the sky.  We evaluate the
noise with which the CIP field can be reconstructed and estimate
the signal-to-noise with which a scale-invariant spectrum of
CIPs may be detected with various experiments.  

We conclude that data from WMAP, Spider, ACT, and SPT are sensitive to CIP amplitudes of $\sim 5-10\%$. The Planck satellite \cite{planck_bluebook} and Polarbear experiment are sensitive to CIP amplitudes as small as $\sim 3\%$. Upcoming ground-based polarization experiments (ACTPol \cite{actpol} and SPTPol \cite{dvorkin_smith_b,sptpol}) or a post-Planck CMB-polarization experiment along the lines of the proposed EPIC experiment \cite{cmbpol} could detect fluctuations of $\sim 0.4\%-0.6\%$. In the cosmic-variance limit, sensitivity to fluctuations of amplitude $\sim 0.05\%$ is possible. 

Our principal motivation in studying CIPs is curiosity: can we
determine empirically, rather simply assume, that the primordial
baryon fraction is homogeneous and traces the dark matter?
Still, there may be theoretical motivation as well.  For
example, curvaton models for inflation may generate CIPs
\cite{lyth_ungarelli_wands,gupta_malik_wands,gordon_lewis_curvaton,enqvist_subdom_curvaton},
with amplitudes approaching the regime detectable by EPIC
\cite{gordon_pritchard}. 
It may also be that recent models
\cite{Kaplan:2009ag,buckley_hylogenesis,cladogenesis,wimp_baryo,heckman_dm,barymorphosis}
that connect the baryon asymmetry and dark-matter density have
implications for CIPs. Additionally, the techniques introduced in this paper could be used to empirically disentangle a CDM isocurvature fluctuation from a baryon isocurvature fluctuation, using CMB data. These modes are usually treated as degenerate in the analysis of CMB observations.

In Ref. \cite{grin_prl}, we presented our basic conclusions. Here we present in detail our results, their derivation, and the computational methods used. We calculate the induced temperature anisotropies in
Sec.~\ref{tempsec} and the induced polarization anisotropies in
Sec.~\ref{polaniso}. In Sec.~\ref{pspec_predict}, we compute the
expected corrections to CMB power spectra for a scale-invariant
spectrum of CIPs and compare the B-mode power spectrum induced
by CIPs at decoupling with that induced at reionization.
In Sec.~\ref{qestsecbig}, we construct minimum-variance
estimators for CIPs.  We then assess in Sec.~\ref{exprop} the
sensitivity of ongoing and upcoming experiments to CIPs, and
we conclude in Sec.~\ref{conc}.  Useful relations involving
tensor spherical harmonics are presented in Appendix \ref{tensor_on_sphere}. Numerical derivatives of transfer
functions are discussed in Appendix
\ref{deriv_pspec_sec}. Second-order harmonic expansions for CMB transfer functions are derived in
Appendix \ref{secdercorr}.  Throughout, we use as our fiducial
cosmological parameters those from Ref.~\cite{wmap7_komatsu_a}.

\section{Perturbed line-of-sight (LOS) formalism: Temperature}
\label{tempsec}

Here we review the standard calculation of the
temperature-fluctuation power spectrum for primordial adiabatic
density perturbations.  We then show how this calculation is altered in the
presence of CIPs.

\subsection{General line-of-sight solution for temperature}

The spherical-harmonic coefficients $T_{lm}$ for the CMB
temperature can be written
\begin{eqnarray}
     T_{lm}&\equiv& \int\,
     d\hat{n}\, T(\hat{n})Y_{lm}^{*}(\hat{n})\nonumber\\
      &=&4\pi\sum_{l_{1}m_{1}} \int\, d\hat{n}\, Y_{lm}^{*}(
      \hat{n}) Y_{l_{1}m_{1}}(\hat{n})
      \int_{0}^{\eta_{0}}\, d\eta\, f(\eta,\hat{n})
      \nonumber \\
      &\times&\int \frac{d^{3}k}{\left(2\pi\right)^{3}}
      \Phi_{\vec{k}}i^{l_{1}} j_{l_{1}}[k
      (\eta_{0}-\eta)]
      Y_{l_{1}m_{1}}^{*}(\hat{k}),
\label{tlosa}  
\end{eqnarray}
where $T\left(\hat{n}\right)$ is the CMB temperature
in direction $\hat{n}$, and $Y_{lm}(\hat n)$ are spherical
harmonics. The Fourier transform of the primordial
gravitational potential for wave-vector $\vec{k}$ is
$\Phi_{\vec{k}}$, while $j_{l}(x)$ denotes a
spherical Bessel function. The conformal time $\eta\equiv \int
dt/a(t)$ is here an integration variable, and $\eta_{0}$ denotes its
value today. The function $f\left(\eta,\hat{n}\right)$, obtained
via the numerical solution of the Boltzmann equations
\cite{cosmics,matias_uros_los,ma_bertschinger},
encodes how much a real-space primordial-potential perturbation
$\Phi[(\eta-\eta_{0})\hat{n}, \eta]$ contributes to the temperature anisotropy
$T(\hat{n})$. It depends on the relation
between initial gravitational-potential fluctuations and
radiation-density fluctuations at decoupling, as well as the recombination history.

\subsection{Temperature anisotropies with homogeneous baryon fraction}

In the standard calculation, this transfer function is the same
in all directions; i.e., $f(\eta,\hat n)= f(\eta)$.  In this
case, Eq.~(\ref{tlosa}) simplifies, via orthogonality of
the $Y_{lm}$s, yielding \cite{matias_harari,matias_uros_los}
\begin{equation}
     T_{lm} = \frac{4\pi i^{l}}{\left(2\pi\right)^{3}} \int
     d\eta \, f(\eta)\int
     d^{3}k\, \Phi_{\vec{k}} j_{l}[k(\eta_{0}-\eta)]
     Y_{lm}^{*}(\hat{k}).
\label{tlosb} 
\end{equation}
The temperature power spectrum is then
easily obtained using Eq.~(\ref{tlosb}), averaging over
realizations of the potential perturbation, and using the
identity $\langle \Phi(\vec{k})\Phi^{*}(\vec{k}')\rangle
= \left(2\pi\right)^{3} \delta_D^{3}(\vec{k}-\vec{k}') P_{\Phi}(k)$,
where $\delta_D^{3}(\vec{k}-\vec{k}')$ is the Dirac
delta function and $P_{\Phi}(k)$ the primordial-potential power
spectrum, and the angle brackets denote an average over
realizations of the primordial potential.
We then find \cite{ma_bertschinger}
\begin{equation}
     \langle T^{*}_{l'm'} T_{lm} \rangle =
     C^{\rm TT}_{l}\delta_{l l'}\delta_{m
     m'},
\label{ClTT}
\end{equation}
where
\begin{equation}
     C_{l}^{\rm TT}=\frac{2}{\pi}\int\, k^{2}\, dk\,
     P_{\Phi}(k)[T_{l}(k)]^{2},
\label{ts1}
\end{equation}
is the CMB temperature power spectrum, written in terms of a
transfer function,
\begin{equation}
     T_{l}(k)=\int\, d\eta\, f(\eta)j_{l}[k(\eta_{0}-\eta)].
\label{t_standard}
\end{equation}
This transfer function is tabulated by Boltzmann codes like
\textsc{camb} \cite{camb} and \textsc{cmbfast}
\cite{matias_uros_los}, and $\delta_{ij}$ is the Kronecker
delta.

\subsection{Temperature anisotropies with CIPs: Single CIP
realization}

In the presence of a compensated isocurvature perturbation, the
baryon and dark-matter fractions vary from one point in the
Universe to another, and so the transfer function $f(\eta,\hat
n)$ now acquires some direction ($\hat n$) dependence.
The CIP involves small changes,
\begin{eqnarray}
     \Omega_{\rm b} \to
     \Omega_{\rm b}\left[1+\Delta\left(\hat{n}\right)\right],\nonumber
     \\ 
     \Omega_{\rm c} \to
     \Omega_{\rm c}-\Omega_{\rm b}\Delta\left(\hat{n}\right),\label{Deltadef}
\end{eqnarray}
in the cosmological parameters between different lines of
sight $\hat n$.  Here, $\Delta(\hat n)$ is the value of the CIP
in direction $\hat n$ at the surface of last scatter (or
reionization---we will make these statements more precise
below).  Note that we define it so that it is the fractional
perturbation in the baryon (rather than dark-matter) density
associated with the CIP.
From Eq.~(\ref{Deltadef}), the change in the total density is $\delta
\Omega_{\rm m}=\delta\Omega_{\rm c}+\delta\Omega_{\rm b}=0$, and so this is
indeed a compensated isocurvature perturbation. 

In a general treatment of perturbed recombination/decoupling, one would follow the set of equations for the electron, dark-matter, photon, and neutrino densities, velocities, and the gravitational potential at second order, as in Refs. \cite{senatore_pertreca,senatore_pertrecb,wandelt_crinkles}. In the case of CIPs, however, the CIP amplitude does not evolve for all observationally accessible scales, and we can thus model the effect of CIPs as a modulation in the cosmological parameters $\Omega_{\rm c}$ and $\Omega_{\rm b}$. 

\begin{figure*}[htbp]
\includegraphics[width=6.50in]{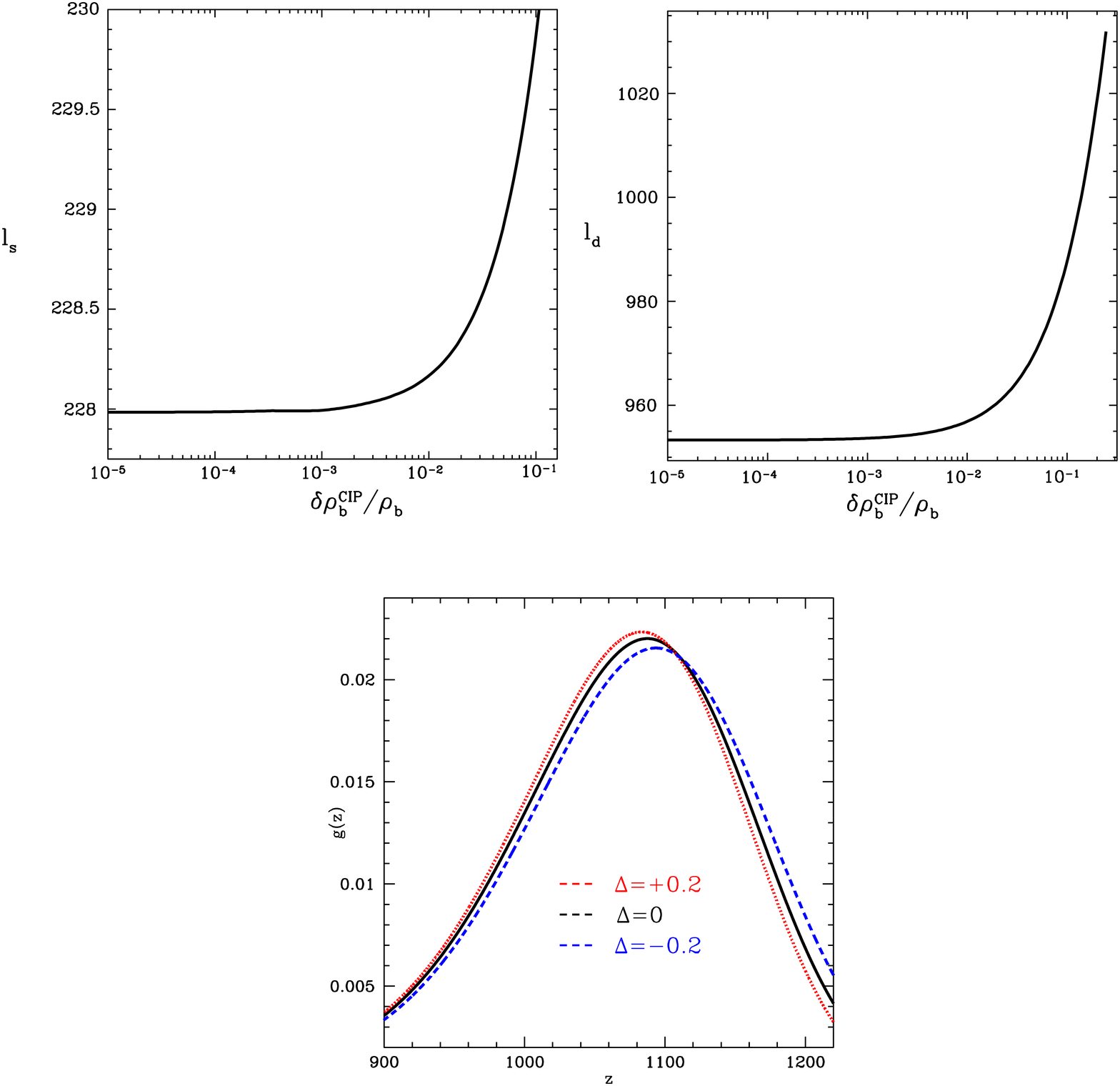}
\caption{Dependence of physically relevant scales and Thomson scattering visibility function $g(z)$ on amplitude of a global CIP perturbation $\Delta$. Top left panel shows dependence of the angular sound horizon $l_{\rm s}$ on $\Delta$. Top right panel shows dependence of the diffusion damping scale $l_{\rm d}$ on $\Delta$. Bottom panel shows $g(z)$ evaluated for $3$ different values of $\Delta$.}
\label{intuit}
\end{figure*}

We can build some intuition for the effect of CIP perturbations on the CIP by considering a globally constant CMB perturbation $\Delta$. We run the \textsc{camb} \cite{camb} code with a global perturbation of the form in Eq.~(\ref{Deltadef}) for a variety of $\Delta$ values. We evaluate the angular sound horizon $l_{s}$ at the surface of last scatter as a function of $\Delta$, using the expressions in Ref. \cite{dodelcosmo}. We see in the top left panel of Fig. \ref{intuit} that, as the plasma is more loaded down with baryons in the presence of a CIP with a positive $\Delta$ value, the decrease in sound speed moves the CMB acoustic peaks to smaller angular scales. 

CMB temperature anisotropies are suppressed on angular scales $l>l_{\rm d}$ due to diffusion damping. Using the expressions in Ref. \cite{matias_harari} and the \textsc{camb} \cite{camb} code, we evaluate $l_{d}(\Delta)$ and show the results in the top right panel of Fig. \ref{intuit}. We see that, as photons diffuse over smaller distances, as a result of higher local baryon density in the presence of a CIP with positive $\Delta$, the transition to exponential damping of CMB anisotropies occurs at higher $l$.

In the bottom panel of Fig. \ref{intuit}, we show the visibility functions $g(z)=e^{-\tau}d\tau/dz$ for $3$ different values of $\Delta$; $\tau$ is the optical depth due to Thomson scattering. The peak of the visibility function is the redshift $z_{\rm SLS}$, at which most CMB photons last scatter. In the presence of a positive (negative) $\Delta$ CIP, decoupling occurs later (earlier) due to higher (lower) baryon density. 

To calculate the effects on the CMB moments $T_{lm}$, we perturb
the line-of-sight solutions, Eqs.~(\ref{tlosa}) and 
(\ref{tlosb}). This approach is relatively simple and
amenable to rapid computation.  The results should be accurate
for multipole moments $L\lesssim 870$ for the CIP, as the baryon
fluctuation can be considered as roughly constant in a given
direction $\hat n$ across the thickness of the surface of last
scatter on such scales.  We discuss in Sec. \ref{pspec_ci}
below how we extrapolate these results to smaller angular
scales ($L\gtrsim 870$) with a Limber approximation.

We proceed by Taylor expanding in real space:
\begin{eqnarray}\nonumber
     f(\eta,\hat{n}) &=& f^{\left(0\right)}
     (\eta) + \Delta(\hat{n}) 
     \frac{df^{\left(0\right)}}{d\Delta} \left(\eta\right)
     \nonumber \\
     &+& \frac{1}{2}\Delta^{2}(\hat{n})
     \frac{d^{2}f^{\left(0\right)}}
     {d^{2}\Delta}\left(\eta\right)+\cdots,
\label{delta_exp} 
\end{eqnarray}
where $f^{\left(0\right)}(\eta)$ is the value of $f$
under the null hypothesis $\Delta(\hat{n})=0$.  We
expand 
\begin{equation}
     \Delta(\hat{n}) =
     \sum_{LM}\Delta_{LM}Y_{LM}(\hat{n}),
\end{equation}
in terms of spherical-harmonic coefficients $\Delta_{LM}$ for
the angular variation in the CIP at the surface of last
scatter.  We then apply the expansion in
Eq.~(\ref{delta_exp}) to linear order in $\Delta$ to the
line-of-sight expression, Eq.~(\ref{tlosa}), and integrate over
angles to obtain the first-order correction,
\begin{align}
     \delta T_{lm}^{\left(1\right)}  = & 4\pi
     \sum_{LM,l_{1}m_{1}} i^{l_{1}} \Delta_{LM}
     \xi^{LM}_{lml_{1}m_{1}} K_{l l_{1}}^{L} \nonumber \\
     \times \int &d\eta \frac{df^{\left(0\right)}}{d\Delta}
     \int \frac{d^{3}k}{\left(2\pi\right)^{3}} \Phi_{\vec{k}}
     j_{l_{1}}[k\left(\eta_{0} -
     \eta\right)]Y_{l_{1}m_{1}}^{*}(\hat{k}),
\label{induced_t}
\end{align}

to $T_{lm}$ in the presence of a CIP, where
\begin{eqnarray}
     \xi^{LM}_{lml_{1}m_{1}} & \equiv &
     {\left(K^{L}_{ll_{1}}\right)}^{-1} \int d\hat{n}Y^{*}_{lm}
     (\hat{n}) Y_{LM}(\hat{n})
     Y_{l_{1}m_{1}} (\hat{n}) \nonumber \\
      &=& \left(-1\right)^{m} \sqrt{\frac{\left(2L+1\right)
      \left(2l+1\right) \left(2l_{1}+1\right)}{4\pi}} \nonumber
      \\ & &\times \wigner{l}{-m}{L}{M}{l_{1}}{m_{1}},\label{xideff}\\
       K^{L}_{ll_{1}}&\equiv&\wigner{l}{0}{L}{0}{l_{1}}{0}, 
\end{eqnarray} 
and the arrays inside parentheses are Wigner-$3J$
symbols.  Throughout, we use the indices $L$ and $M$ exclusively
for the decomposition of the CIP, while lower-case
indices are used for the multipole moments of the CMB
observables $T_{lm}$, $E_{lm}$, and $B_{lm}$. Sums over $m$ ($M$)
are always taken over the range
$-l\leq m\leq l$ ($-L\leq M\leq L)$, while sums over $l$ ($L$)
are formally taken over $1\leq l\leq \infty$ ($1\leq L\leq
\infty$); in practice, a maximum value $l_{\rm max}$ is used for
numerical evaluation, as discussed in
Sec.~\ref{pspec_predict}. The monopole $L=0$ corresponds to a
global shift in $\Omega_{b}$ and $\Omega_{c}$ and we absorb this
term into the cosmological parameters themselves.
  
For a given realization of the CIP field---that is, for a given
set of $\Delta_{LM}$---the covariance between temperature
moments is now
\begin{equation}
     \langle T^{*}_{l'm'}T_{lm} \rangle\simeq
     C_{l}^{\rm
     TT,\left(0\right)}+\sum_{LM}\Delta_{LM}\xi^{LM}_{lml'm'}S_{l
     l'}^{L,{\rm TT}},
\label{toffdia_a}
\end{equation}
where
\begin{equation}
     S^{L,{\rm TT}}_{ll'} \equiv
     \left(C_{l'}^{\rm T,dT}+C_{l}^{\rm
     T,dT}\right)K_{ll'}^{L},
\label{toffdia_middle}
\end{equation}
and
\begin{equation}
     C_{l}^{\rm X, dX'} \equiv \frac{2}{\pi} \int k^{2}\,
     dk\, P_{\Phi}(k) X_{l}(k) \frac{dX'_{l}(k)}{d\Delta},
\label{toffdia_b}
\end{equation}
for $\{ \rmX,\rmX' \} \in \{ \rmT,\rmE,\rmB \}$ (a
generalization that will be useful below), and
$C_{l}^{{\rm TT}, \left(0\right)}$ is the temperature
power spectrum in the absence of CIPs.  Here,
\begin{equation}
     \frac{dX'_{l}(k)}{d\Delta} = \int d\eta \frac{
     df(\eta)}{d\Delta} j_{l}[k(\eta_{0}-\eta)]
\label{firstderiv}
\end{equation}
describes the change in the transfer function,
Eq.~(\ref{t_standard}), with $\Delta$.

In deriving these results, we have taken an average over
realizations of the primordial-potential power spectrum
$\Phi_{\vec{k}}$, but we have restricted our consideration to a
given realization of the CIP.  In
Sec.~\ref{qestsecbig}, we build the formalism to reconstruct
$\Delta(\hat{n})$ from these off-diagonal temperature
correlations as well their generalization to polarization.

\subsection{Temperature anisotropies with CIPs: Average over CIP
realizations}
\label{taniso_ensemble}

We now take an ensemble average
over many realizations of both the primordial potential field
and the CIP field. This allows us, given a spectrum of CIPs, to
calculate the effects of these CIPs on the power spectrum of CMB
fluctuations measured on the entire sky. 

We denote the ensemble average  of a spatially varying field $X$
over realizations of the CIP field by $\langle X 
\rangle_{\rm b}$. We denote the ensemble average over
realizations of both the CIP field and the primordial potential
by $\langle X \rangle_{\rm bc}$. From
Eqs.~(\ref{toffdia_a}) and (\ref{toffdia_middle}), we see that
$\langle T^{*}_{l'm'}T_{lm}^{} \rangle_{\rm
bc}\propto \langle \Delta_{LM}\rangle_{\rm b}$. For an isotropic
random field, $ \langle \Delta_{LM}\rangle_{\rm b}=0$, so we
must thus go to second order in $\Delta(\hat{n})$ to
compute the effects of CIPs on the CMB power spectrum.
We thus obtain, to second order in $\Delta$, the temperature
power spectrum,
\begin{align}
     C_{l}^{\rm TT,(2)}&\equiv  \VEV{|T_{lm}|^2}_{\rm bc}
     \simeq C_{l}^{\rm
     TT,\left(0\right)} + \left\langle \left|\delta
     T^{\left(1\right)}_{lm} \right|^2 \right\rangle_{\rm
     bc}\nonumber \\
     &+\left\langle
     T^{\left(0\right)*}_{lm}\delta
     T_{lm}^{\left(2\right)}\right\rangle_{\rm bc}+
     \left\langle \delta T^{\left(2\right)*}_{lm} 
     T_{lm}^{\left(0\right)}\right\rangle_{\rm bc},
\label{t2los_a} 
\end{align} 
where $T_{lm}^{\left(0\right)}$ is the unperturbed
moment in Eq.~(\ref{tlosb}), $\delta
T_{lm}^{\left(1\right)}$ is given by Eq.~(\ref{induced_t}), and
$C_{l}^{\rm TT,\left(0\right)}$ is the unperturbed power
spectrum given by Eq.~(\ref{ts1}). The $^{(2)}$ superscript
denotes the term arising when expanding $T_{lm}$ to order
$\Delta^{2}(\hat{n})$.  We evaluate this term using the
second-derivative term $d^{2}f^{\left(0\right)}/d\Delta^{2}$ in
Eq.~(\ref{harmonic}). We take an expectation value over CIPs and
primordial-potential realizations. We then use
Eqs.~(\ref{tlosa}), (\ref{xideff}), and (\ref{t2los_a}), identities of Wigner-$3J$ symbols \cite{angmom}, and
Appendix~\ref{secdercorr} to obtain
\begin{eqnarray}
     C_{l}^{\rm TT,\left(2\right)} & = & C_{l}^{\rm
     TT,\left(0\right)} + \delta C_{l}^{\rm TT,\left(1\right)} +
     \delta C_{l}^{\rm TT,\left(2\right)} \nonumber,\\
     \delta C_{l}^{\rm TT,\left(1\right)} & \equiv &
     \sum_{L,l_{1}} C_{L}^\Delta C_{l_{1}}^{\rm dT,dT} 
     \left(K_{l l_{1}}^{L}\right)^{2} G_{Ll_{1}},
     \nonumber\label{t_firstorder}\\
     G_{Ll_{1}} & \equiv & \left[\frac{\left(2L+1\right)
     \left(2l_{1}+1\right)} {4\pi}\right],\\
     \delta C_{l}^{\left(2\right)} &\equiv& \Delta_{\rm
     bc}^{2}C_{l}^{\rm T,d^{2}T}.\label{tpspec_res}
\end{eqnarray}
The CIP power spectrum $C_{L}^\Delta$ and total variance
$\Delta_{\rm bc}^{2}$ obey
\begin{align}
     \left\langle |\Delta_{LM}|^2\right\rangle_{b} & \equiv
     C_{L}^\Delta,\label{pspec_bc}\\
     \Delta_{\rm
     bc}^{2}&=\sum_{L}\left(\frac{2L+1}{4\pi}\right)C_{L}^\Delta,
     \label{normalization_cl}
\end{align}
while the CMB derivative power spectra are given by 
\begin{align}
     C_{l}^{\rm dX,\rm dX'}&=&\frac{2}{\pi}\int k^{2}\, dk\,
     P_{\Phi}(k)\frac{dX_{l}(k)}{d\Delta}\frac{dX'_{l}(k)}{d\Delta},\label{derpspec_computenot}
\\
     C_{l}^{\rm X, \rm d^{2}X'}&=&\frac{2}{\pi}\int
     k^{2}\, dk\, P_{\Phi}(k)X_{l}(k)
     \frac{d^{2}X'_{l}(k) } {
     d\Delta^{2}},
\label{derpspec_compute}
\end{align}
where $d^2X_l/d^2\Delta$ are defined analogously to the
first-derivative transfer function in Eq.~(\ref{firstderiv}).
Appendix \ref{deriv_pspec_sec} details the calculation of
the derivative power spectra.

\section{Perturbed line-of-sight formalism: Polarization}
\label{polaniso}

We now generalize the analysis above to the CMB polarization.
In addition to inducing off-diagonal
correlations in the polarization spherical-harmonic coefficients,
CIPs will induce B modes. We begin by
reviewing the LOS solution for polarization under the null
hypothesis of no CIPs. We then compute the
effects of CIPs, both for a single realization of the CIPs, and then for an average over realizations of a spectrum
of CIPs. 

\subsection{Polarization anisotropies with homogeneous baryon fraction}

Polarization is a spin-2 tensor field and can be expanded as
\cite{kamion_polarization,matias_uros_polarization}
\begin{eqnarray}
     P_{ab}\left(\hat{n}\right)=\sum_{lm}
     \left(E_{lm}Y_{lm,ab}^{\rm E}+B_{lm}Y_{lm,ab}^{\rm
     B}\right),
\label{ptens}
\end{eqnarray} 
where $Y_{lm,ab}^{\rm E}$ and $Y_{lm,ab}^{\rm B}$ are the E- and
B mode (``grad" and ``curl", respectively) tensor spherical harmonics, as
defined in Appendix~\ref{tensor_on_sphere}. The right-most
indices after the comma are tensor indices. In terms of the
Stokes polarization parameters $Q$ and $U$, the
polarization tensor is
\cite{kamion_polarization,matias_uros_polarization}
\begin{equation}
     P_{ab}\left(\hat{n}\right) = \frac{1}{2}
     \left( \begin{array}{cc}  Q \left(\hat{n}\right) & -
     U\left(\hat{n}\right)\sin{\theta}\\-
     U\left(\hat{n}\right)\sin{\theta}&-
     Q\left(\hat{n}\right)\sin^{2}\theta\end{array} 
    \right),
\end{equation}
where $\theta$ is the polar angle of the LOS with respect to some origin.

Under the null hypothesis, the polarization pattern at the
surface of last scatter is a pure E mode, with multipole moments
given by 
\begin{equation}
     E_{lm}=\frac{4\pi i^{l}}{\left(2\pi\right)^{3}}\int d\eta\,
     f^{\rm E}(\eta)\int
     d^{3}k\Phi_{\vec{k}}j_{l}(x)Y_{lm}^{*}(\hat{k}), 
\label{eusual}
\end{equation}
where $x\equiv k(\eta_{0}-\eta)$, and $f^{E}(\eta)$ is
the E-mode transfer function, obtainable numerically from
Boltzmann codes.  The polarization covariance and TE covariance
are derived analogously to the results for temperature, yielding
\cite{matias_uros_los} 
\begin{eqnarray}
     \langle E^{*}_{l'm'}E_{lm} \rangle=C^{\rm
     EE}_{l}\delta_{l l'}\delta_{m
     m'},     \nonumber\\ 
     C_{l}^{\rm EE} = \frac{2}{\pi} \int k^{2} dk P_{\Phi}(k)
     [E_{l}(k)]^{2},\nonumber\label{es1}\\ 
     E_{l}(k) = \int d\eta f^{\rm E}(\eta)
     j_{l}[k(\eta_{0}-\eta)], \label{e_standard}
\end{eqnarray}
and \cite{matias_uros_los}
\begin{eqnarray}
     \langle E^{*}_{l'm'}T_{lm} \rangle=C^{\rm
     TE}_{l}\delta_{l l'}\delta_{m
     m'},\nonumber\\
     C_{l}^{\rm TE}=\frac{2}{\pi}\int k^{2} dk
     P_{\Phi}(k)T_{l}(k)E_{l}(k)\label{testandard}.
\end{eqnarray}

\subsection{Polarization anisotropies with CIPs: single CIP realization}

We now generalize the analysis to include the effects of CIPs.
In the presence of a CIP field $\Delta(\hat n)$, the real-space
polarization tensor may be Taylor expanded as
\begin{eqnarray}
     P_{ab}(\hat{n}) & = & P_{ab}^{\left(0\right)}
     (\hat{n}) + \frac{ dP_{ab}^{\left(0\right)}}
     {d\Delta} (\hat{n}) \Delta( \hat{n})
     \nonumber\\
     &+&\frac{1}{2}\frac{d^{2}P_{ab}^{\left(0\right)}}
     {d\Delta^{2}} \left(\hat{n}\right) \Delta^{2}
     (\hat{n}) +\cdots.
\label{polarization_taylor}
\end{eqnarray}
Just as in the case of temperature, when considering a single
realization, we need only consider the first-order terms in
Eq.~(\ref{polarization_taylor}). We then utilize
Eqs.~(\ref{ptens}), (\ref{eusual}), and the
first-derivative piece of the usual expansion for
$\Delta(\hat{n})$ [see Eq.~(\ref{delta_exp})] to obtain an
expansion for the polarization tensor $P_{ab}$ in the presence
of CIPs:
\begin{eqnarray}
     P_{ab}\left({\hat
     n}\right)&=&P_{ab}|^{\Delta=0}+\left.\delta
     P_{ab}\right|^{1}+\left.\delta P_{ab}\right|^{2}+\cdots,\\
     \delta P_{ab}^{\left(1\right)}&\equiv&
     \sum_{l_{1}m_{1}} \frac{dE_{l_{1}m_{1}}}
     {d\Delta}Y_{l_{1}m_{1},ab}^{E}(\hat{n}) \Delta
     \left(\hat{n}\right) \nonumber\\
     &=&\sum_{l_{1}m_{1}}^{LM}\frac{dE_{l_{1}m_{1}}}{d\Delta}Y_{l_{1}m_{1}
     ab}^{E}(\hat{n})\Delta_{LM}Y_{LM}\left(\hat{n}\right). 
\end{eqnarray}
We may now pick off the induced E- and B-mode multipole moments
$\delta E_{lm}^{\left(1\right)}$ and $\delta
B_{lm}^{\left(1\right)}$ at order $\Delta$, using the appropriate
integral over a tensor spherical harmonic:
\begin{eqnarray}
     \delta E_{lm}^{\left(1\right)} = \int d\hat{n}\,
     Y_{lm,ab}^{{\rm E},*}\left(\hat{n}\right) \delta
     P_{ab}^{\left(1\right)},\label{intdefa}\\
     B_{lm}=\delta B_{lm}^{\left(1\right)}=\int d\hat{n}\,
     Y_{lm,ab}^{{\rm B},*}\left(\hat{n}\right) \delta
     P_{ab}^{\left(1\right)}.
\label{intdefb} 
\end{eqnarray}
We evaluate Eqs.~(\ref{intdefa})--(\ref{intdefb}), calling on
Eqs.~(\ref{epair})--(\ref{spinint}),
yielding 
\begin{align}
     \delta E_{lm}^{\left(1\right)} & =
     \sum_{LM,l_{1}m_{1}}^{L+l_{1}+l~{\rm even}}
     \xi^{LM}_{lm,l_{1}m_{1}}H^{L}_{ll_{1}}\Delta_{LM}
     \frac{dE_{l_{1}m_{1}}}{d\Delta},\label{ei}\\
     \delta
     B_{lm}^{\left(1\right)}&=\sum_{LM,l_{1}m_{1}}^{L+l_{1}+l~{\rm
     odd}}\left(-i\right)
     \xi^{LM}_{lm,l_{1}m_{1}}H^{L}_{ll_{1}}\Delta_{LM}
     \frac{dE_{l_{1}m_{1}}}{d\Delta},\label{bi}
\end{align}
where
\begin{equation}
     H^{L}_{ll_{1}}\equiv \wigner{l}{2}{L}{0}{l_{1}}{-2}.
\end{equation}
We now evaluate the induced correlations between different
temperature/polarization moments. At first order in
$\Delta(\hat{n})$, $\langle
B^{*}_{l'm'}B_{lm}\rangle\propto \langle
B^{*\left(0\right)}_{l'm'}\delta
B_{lm}^{\left(1\right)}\rangle=0$. The remaining covariances
are 
\begin{align}
     \langle E_{l'm'}^{*}
     E_{lm}\rangle=&~C_{l}^{\rm EE}\delta_{l
     l'}\delta_{m m'}\nonumber\\ 
     +&\sum_{LM}^{L+l+l'~{\rm
     even}}\Delta_{LM}\xi^{LM}_{lm,l'm'}S^{L,{\rm
     EE}}_{l l'},\nonumber\\ 
     S_{l l'}^{L,{\rm EE}}\equiv&\left(C_{l}^{\rm E,
     dE}+C_{l'}^{\rm E, dE}\right)H^{L}_{l
     l'},\label{induced_ee_offdiag}\\
     \langle E_{l'm'}^{*}
     B_{lm}\rangle=&\sum_{LM}^{L+l+l'~{\rm
     odd}}\Delta_{LM}\xi^{LM}_{lm,l'm'}S^{L,{\rm
     EB}}_{l l'},\nonumber\\
     S_{l l'}^{L,{\rm EB}}\equiv&-iC_{l'}^{\rm
     E, dE}H^{L}_{l l'},\\
     \langle T_{l'm'}^{*}
     B_{lm}\rangle=&\sum_{LM}^{L+l+l'~{\rm
     odd}}\Delta_{LM}\xi^{LM}_{lm,l'm'}S^{L,{\rm
     TB}}_{l l'},\nonumber\\
     S_{l l'}^{L,\rm TB}\equiv&-iC_{l'}^{\rm T,
     dE}H^{L}_{l l'},\\
     \langle T_{l'm'}^{*}
     E_{lm}\rangle=&~C_{l}^{\rm TE}\delta_{l
     l'}\delta_{m m'}\nonumber\\
     +&\sum_{LM}^{L+l+l'~{\rm
     even}}\Delta_{LM}\xi^{LM}_{lm,l'm'}S_{l
     l'}^{L,{\rm TE}}\nonumber,\\
     S_{l l'}^{L,{\rm
     TE}}\equiv&~\left(C_{l'}^{\rm T,dE}H^{L}_{l
     l'}+C_{l}^{\rm E,dT}K^{L}_{l
     l'}\right).\label{induced_te_offdiag}
\end{align}

\subsection{Polarization anisotropies with CIPs: Average over
CIP realizations}
\label{polaniso_ave}

We now extend the ensemble average to multiple realizations of
the CIP field.  We do this in order to compare
the polarization power spectrum induced by CIPs at the surface of last scatter
with that induced at reionization.
For temperature, the average over realizations of both the CIP
and primordial-potential perturbations is given by
Eq.~(\ref{t2los_a}). Extending this average to
$X,X'\in\left\{T,E,B\right\}$, we obtain the XX$'$ power
spectra, averaged over the entire sky, to second order in $\Delta$:
\begin{align}
     C_{l}^{{\rm X X'},\left(2\right)} &\equiv\left\langle
  X_{lm} X_{lm}^{'*}\right\rangle_{\rm bc} \nonumber \\
     &\simeq C_{l}^{\rm
     XX',\left(0\right)} +\left\langle 
     \delta X^{\left(1\right)}_{lm}\delta X_{lm}^{\prime\left(1\right)} \right\rangle_{\rm
     bc}\nonumber\\
     &+\left\langle
     X^{\left(0\right)*}_{lm}\delta
     X_{lm}^{\prime\left(2\right)}\right\rangle_{\rm bc}+
     \left\langle \delta
     X^{\left(2\right)*}_{lm} 
     X_{lm}^{\prime\left(0\right)}\right\rangle_{\rm bc},
\label{general_perturbed_t}
\end{align}
where $C_{l}^{{\rm X X^{\prime}},(0)}$ is the power spectrum computed with no CIP contribution.

We evaluate Eq.~(\ref{general_perturbed_t}) with
Eqs.~(\ref{spinint}) and (\ref{4spin}) and Wigner-$3J$ relations to
simplify the resulting integrals and sums. Superscript indices
$^{\left(1\right)}$ and $^{\left (2 \right)}$ indicate the order
of the derivative $d^{n}f^{\left(0\right)}/d\Delta^{n}$ used to
derive the indicated term, as in Sec.~\ref{taniso_ensemble}. The
resulting nonzero power spectra are
\begin{align}
     C_{l}^{\rm TE}\simeq&~~C_{l}^{{\rm
     TE},\left(0\right)}+\delta C_{l}^{{\rm
     TE},\left(1\right)}+\delta C_{l}^{{\rm
     TE},\left(2\right)}\label{te_pspec_new}\\
     \delta C_{l}^{{\rm
     TE},\left(1\right)}\equiv&~~\sum_{L,l_{1}}^{L+l_{1}+l~{\rm
     even}}C_{L}^\Delta C_{l_{1}}^{\rm d{ T}, d{\rm
     E}}G_{Ll_{1}}H_{l l_{1}}^{L}K_{l l_{1}}^{L}\nonumber\\
     \delta C_{l}^{{\rm
     TE},\left(2\right)}\equiv&~\frac{\Delta_{\rm
     bc}^{2}}{2}\left(C_{l}^{{\rm T},{\rm d}^{2}{\rm
     E}}+C_{l}^{{\rm E},{\rm d}^{2}{\rm T}}\right),\nonumber
\end{align}
\begin{align}
     C_{l}^{\rm EE}\simeq&~~C_{l}^{{\rm
     EE},\left(0\right)}+\delta C_{l}^{{\rm
     EE},\left(1\right)}+\delta C_{l}^{{\rm
     EE},\left(2\right)},\label{ee_pspec_new}\\
     \delta C_{l}^{{\rm
     EE},\left(1\right)}\equiv&~~\sum^{L+l_{1}+l~{\rm
     even}}_{L,l_{1}}C_{L}^\Delta C_{l_{1}}^{ {\rm dE},  {\rm
     dE}}G_{Ll_{1}}\left(H_{l l_{1}}^{L}\right)^{2},\nonumber\\
     \delta C_{l}^{{\rm EE},\left(2\right)}=&~~\Delta_{\rm
     bc}^{2}C_{l}^{{\rm E},{\rm d}^{2}{\rm E}},\nonumber
\end{align}
and 
\begin{align}
     C_{l}^{\rm BB}\simeq&~~\sum^{L+l_{1}+l~{\rm
     odd}}_{L,l_{1}}C_{L}^\Delta C_{l_{1}}^{{\rm dE}, {\rm
     dE}}G_{Ll_{1}}\left(H_{l
     l_{1}}^{L}\right)^{2}.
\label{bb_pspec_new}
\end{align}
The CIP field $\Delta(\hat n)$ is a scalar and cannot
statistically change the parity of polarization perturbations.
This requires that $C_{l}^{\rm TB}$ and $C_{l}^{\rm EB}$ vanish when
averaging over CIP realizations.  Algebraically, this is
enforced by the vanishing of the relevant Wigner-$3J$ symbols,
as occurs with optical-depth fluctuations at reionization
\cite{dvorkin_smith_a,dvorkin_smith_b} and with
gravitational-potential perturbations in weak lensing
\cite{wayne_harmonic,takemi_wayne_c}. Indeed, the geometric (Wigner-$3J$) symbols obtained are the same as for those effects.  CIPs give rise to different $ll'$ dependences for the functions $S_{l
l'}^{L,{\rm X X'}}$, however, through the dependence on the derivative power spectra $C_{l}^{\rm X X'}$, allowing them to be disentangled observationally from gravitational-potential fluctuations along the LOS or optical-depth fluctuations at reionization.

\section{Numerical Results for B- Mode Power Spectra}
\label{pspec_predict}

We now apply the formulas derived in Secs.~\ref{taniso_ensemble}
and \ref{polaniso_ave} to compute the power spectra for B modes
induced by CIPs at decoupling.  We first discuss the form of
the angular CIP power spectrum $C_{L}^\Delta$.  We then present
numerical results for B modes induced at decoupling.  For
comparison, we then reproduce the calculations of
Ref.~\cite{holder} of the B modes induced at reionization.

\subsection{Power spectrum of compensated perturbations}
\label{pspec_ci}

\subsubsection{Three-dimensional CIP power spectrum}

To proceed further, we must make an \textit{ansatz} for the
spectrum of CIPs. Motivated by the curvaton model
(which produces a nearly scale-invariant spectrum of CIPs)
\cite{lyth_ungarelli_wands,gordon_lewis_curvaton,gordon_pritchard},
we assume a scale-invariant spectrum for the three-dimensional
CIP field $\Delta(\bf x)$; that is,
\begin{align}
      \left \langle \tilde
      \Delta^*(\mathbf{k}')\tilde \Delta
      (\mathbf{k})\right \rangle=& (2\pi)^3 \delta_D^{3}
      \left(\mathbf{k}-\mathbf{k}'\right) P_\Delta(k) 
      \nonumber\\ 
      P_\Delta(k)=&~Ak^{-3},
\label{pci}
\end{align}
where $\tilde \Delta(\mathbf{k})$ is the Fourier transform of
$\Delta(\bf x)$, and $A$ is a dimensionless CIP amplitude.  

As discussed in the introduction, the strongest constraint to
$A$ comes from cluster baryon fractions.  This constraint tells
us that the variance,
\begin{equation}
     \Delta^2_{\rm cl} = \frac{1}{2\pi^2} \int k^2 \,
     dk\, \left[ 3 j_1(kR)/(kR) \right]^2 P_\Delta(k),
\label{eqn:clusterrms}
\end{equation}
in the baryon--to--dark-matter ratio on $R\sim10$~Mpc scales is
$\Delta_{\rm cl} \lesssim 0.08$.  The integral has a formal
logarithmic divergence at low $k$ which is cut off, however, by
the volume occupied by the clusters surveyed.  Taking this to be
the horizon, $k_{\mathrm{min}}\simeq (10~\mathrm{Gpc})^{-1}$, we
find $A\lesssim0.017$. Since the cosmological baryon fraction $\Omega_{\rm b}$ determines primordial abundances via BBN, there is an additional constraint from astrophysical measurements of these abundances \cite{holder}. However, this constraint is less stringent than the one from cluster gas fractions.

\label{clusterbound}

\subsubsection{Angular CIP power spectrum}
\label{angularcipps}
When the 3-dimensional field is projected onto a narrow
spherical surface, the resulting angular power spectrum for
$\Delta$ will be $C_L^\Delta\simeq A/(\pi L^2)$ for mulipole moments
$L\lesssim (\eta_0-\eta_{\mathrm{ls}})/\sigma_\eta \simeq 870$, where
$\eta_{\mathrm{ls}}$ and $\eta_0$ are the conformal time at last
scatter and today, respectively, and $\sigma_\eta$ is the rms
conformal-time width of the surface of last scattering (SLS).  At smaller
angular scales (larger $L$), the angular variation in $\Delta$
is suppressed by the finite width \cite{silkdamping} of the scattering surface.
Using the Limber approximation, the angular power spectrum for
$\Delta$ can be approximated by $C_L^\Delta \simeq A
(\eta_0-\eta_{\mathrm{ls}})/(2\sqrt{\pi} L^3\sigma_\eta)$ for
$L\gtrsim 870$.  The
exact analytic expression we use is obtained from the Limber
approximation, approximating the visibility function as a
Gaussian. It is
\begin{equation}
     C_L^\Delta =\frac{A}{2\sqrt{\pi}
     L^{2}}U\left[\frac{1}{2},0,\left(\frac{L\sigma_{\eta}}{\eta_{0}-\eta_{\rm
     sls}}\right)^{2}\right],
\label{precise_cL}
\end{equation}
where $U(a,b,x)$ is a confluent hypergeometric function.  We use
$\eta_{0}-\eta_{\rm ls}=14100~{\rm Mpc}$ and $\sigma_\eta=16.2~{\rm Mpc}$ for decoupling. We use
$\eta_{0}-\eta_{\rm ls}=9760~{\rm Mpc}$ and $\sigma_\eta=448~{\rm Mpc}$ for reionization. These values are obtained by directly fitting to the visibility function output by the \textsc{camb} code \cite{camb}. Of course, Eq.~(\ref{precise_cL}) is an approximation, and the precise shape of the transition from $C_{L}^\Delta \propto 1/L^{2}\to C_{L}^{\Delta}\propto 1/L^{3}$ near $L\sim 870$ depends on the interference of Fourier modes of $\Delta$ with those of $\Phi$, averaged over the SLS. This issue warrants future study, but the asymptotic behavior at low and high $L$ is correct (as shown for an analogous computation in Ref. \cite{levon_magnet}). Moreover, as we shall see in Sec. \ref{exprop}, most signal-to-noise in CIP reconstruction comes either from $L\lesssim100$ or $L\gsim 2000$, and so the main conclusions of this work should not be affected.

\begin{figure}[htbp]
\includegraphics[width=3.26in]{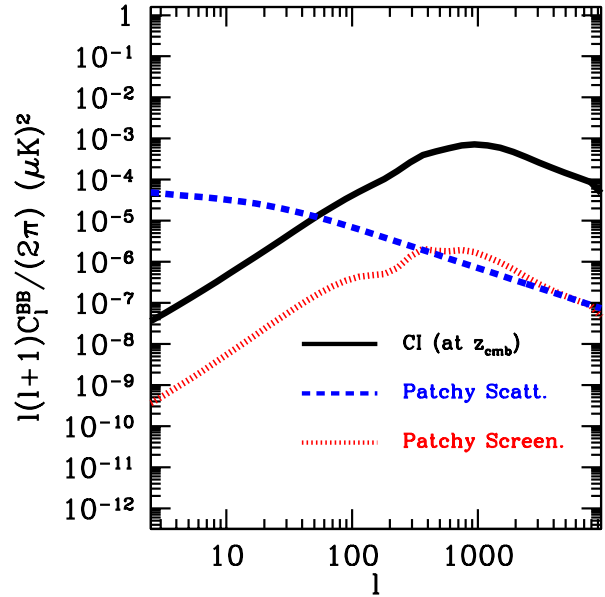}
\caption{CMB B-mode polarization power spectra induced by CIP
     perturbations at decoupling (black solid line), compared
     with the effects of CIPs at reionization, for which two
     contributions are shown: patchy screening (red dotted
     line), and patchy scattering (blue short-dashed line). The amplitude for the CIP power
     spectrum is that which saturates the $\Delta_{\rm cl}
     \lesssim 0.08$ bound from clusters \cite{holder}.}
\label{dvork}
\end{figure}

\subsection{Numerical result for B modes from CIPs at
decoupling}

Using the Limber approximation with values
$\eta_{0}-\eta_{\mathrm{ls}}=14100~{\rm Mpc}$ and $\sigma_\eta=16.2~{\rm Mpc}$ appropriate for decoupling, Eqs.~(\ref{te_pspec_new})--(\ref{bb_pspec_new})
can be used to obtain predictions for the B modes induced by
CIPs at decoupling.  The results are shown in
Fig.~\ref{dvork} for $A=0.017$, the largest CIP amplitude
consistent with the galaxy-cluster constraint. Appendix \ref{deriv_pspec_sec} details the calculation of the requisite derivative power spectra. We use a maximum $l$ value of $l_{\rm max}=10000$.
 
\subsection{Reionization}
\label{patchy_reion_sec}

In Ref.~\cite{holder}, it was noted that spatial inhomogeneities
in the baryon density give rise to angular variations in
the optical depth $\tau$ for rescattering of CMB photons at
reionization. It was also noted that these inhomogeneities
would give rise to B modes primarily at large angular scales by
patchy rescattering of CMB photons and at smaller angular
scales through patchy screening of the primary CMB polarization
from the decoupling epoch.  These calculations build upon
calculations in
Refs.~\cite{limbera,Baumann:2002es,dore_patchy,dvorkin_smith_a,dvorkin_smith_b} where
optical-depth fluctuations were postulated to arise from
inhomogeneities in the free-electron fraction due to
inhomogeneous reionization.

In our notation, the contribution of patchy screening is
\cite{dvorkin_smith_a,dvorkin_smith_b}
\begin{align}
     C_{l}^{\rm TT}=& \tau^2
     e^{-2\tau}\sum_{L,l_{1}}C_{L}^{\Delta}C_{l_{1}}^{\rm
     TT,rec}\left(K_{l
     l_{1}}^{L}\right)^{2}G_{Ll_{1}}\\
     \delta C_{l}^{\rm
     TE}=& \tau^2 e^{-2\tau}\sum_{L,l_{1}}^{L+l_1+l~{\rm
     even}}C_{L}^{\Delta}C_{l_{1}}^{\rm TE,rec}K_{l
     l_{1}}^{L}H_{l l_{1}}^{L}G_{Ll_{1}},\\
     \delta C_{l}^{\rm
     EE}=& \tau^2 e^{-2\tau}\sum_{L,l_{1}}^{L+l_1+l~{\rm
     even}}C_{L}^{\Delta}C_{l_{1}}^{\rm EE,rec}\left(H_{l
     l_{1}}^{L}\right)^{2}G_{Ll_{1}},\\
     \delta C_{l}^{\rm
     BB}=& \tau^2 e^{-2\tau}\sum_{L,l_{1}}^{L+l_1+l~{\rm
     odd}}C_{L}^{\Delta}C_{l_{1}}^{\rm EE,rec}\left(H_{l
     l_{1}}^{L}\right)^{2}G_{Ll_{1}},
\label{patchy_screen_eq}
\end{align}
where $\tau=0.086$ is the mean optical depth, and $C_L^\Delta$
is here the angular CIP power spectrum {\it for reionization};
i.e., obtained with $\eta_{0}-\eta_{\mathrm{ls}}=9760~{\rm Mpc}$ and $\sigma_\eta=448~{\rm Mpc}$. These values are obtained by fitting a Gaussian visibility function to the reionization model of Ref. \cite{lewis_reion}.

The contributions of patchy scattering are
\begin{align}
     \delta C_{l}^{\rm BB} = \delta C_{l}^{\rm EE}=\frac{3\tau^2
     }{100}C_{l}^{\Delta}Q_{\rm rms}^{2}e^{-2\tau},
\label{patchy_scatt_eq}
\end{align}
where $Q_{\rm rms}\simeq 17.9\, \mu{\rm K}$ is the rms
temperature quadrupole at reionization.

Figure \ref{dvork} shows the B modes induced by patchy
scattering and screening at reionization again using $A=0.017$.
We see that at all but the largest scales, the
decoupling-induced B modes are larger (by up to $\sim 3$
orders of magnitude) than those induced at reionization. 

We thus conclude that the effects of CIPs on CMB fluctuations
would be much larger than found in Ref.~\cite{holder},
particularly at the small scales most important for detection
and reconstruction  (to be discussed below) of CIPs from the CMB.   
We thus now move on to show how spatial fluctuations in the
baryon--to--dark-matter ratio can be measured with CMB maps.

\section{Measurement of CIPs with the CMB}
\label{qestsecbig}

In this section, we show how the CIP field $\Delta(\hat n)$ can
be measured with off-diagonal CMB correlations, building upon
analogous prior work on measurement of cosmic-shear fields
\cite{uros_cmb_lens_a,uros_cmb_lens_b,zaldar_lens,uros_evolve,wayne_harmonic,takemi_wayne_c},
departures from statistical isotropy
\cite{souradeep_a,souradeep_b,pullen_kamion}, and cosmic
birefringence
\cite{kamion_derotate,gluscevic_kamion_fullsky,yadav_rotate,gluscald}.
Having concluded that the decoupling signal is much bigger
than that from reionization, we consider detection/measurement
of  CIPs at the surface of last scatter.

In Sec.~\ref{mvdelta_sec}, we construct a minimum-variance
quadratic estimator $\widehat{\Delta}_{LM}$ for the multipole
moments of the CIP field. In Sec.~\ref{Dcovar_sec}, we
explicitly calculate the noise covariance (due both to cosmic
variance and experimental noise) needed to evaluate the errors
and optimal weights of Sec.~\ref{mvdelta_sec}. Finally, in
Sec.~\ref{SN_sec}, we use the results of the preceding sections
to derive an expression for the signal-to-noise ratio (SNR) with
which a CMB experiment can detect CIPs. 

\subsection{Minimum-variance estimators for $\Delta_{LM}$}
\label{mvdelta_sec}

The total correlation between multipole moments (including the contribution induced by a given realization of CIPs)
takes the form [see Eq.~(\ref{toffdia_a}) and
(\ref{induced_ee_offdiag})--(\ref{induced_te_offdiag})],
\begin{align}
     \left \langle X_{l'm'}^{'*}
     X_{lm}\right\rangle = \nonumber& \\C_{l}^{\rm X X'} \delta_{l
     l'}\delta_{m m'}&+\sum_{LM}D^{LM,{\rm X
     X'}}_{l l'}\xi^{LM}_{lm,l'm'},\label{correlaa}
\end{align}
where
\begin{equation}
     D^{LM,{\rm X X'}}_{ll'}\equiv \Delta_{LM}
     S_{l l'}^{L,{\rm X
     X'}}\nonumber.\label{correlab}
\end{equation} 
As before, ${\rm X, X'}\in \left\{{\rm T,E,B}\right\}$.
As discussed in Secs.~\ref{tempsec} and \ref{polaniso}, the
functions $S_{l l'}^{L,{\rm X X'}}$ map
$\Delta_{LM}$ to observed off-diagonal CMB anisotropies. 
The $S_{l l'}^{LM,{\rm X X'}}$ are assembled in
Table~\ref{mfunc}.  

\begin{table}
\caption{The ``response functions" $ S_{l l'}^{L,{\rm X
      X'}}$ of CMB fluctuations to CIPs, defined
      in Eqs.~(\ref{toffdia_a}) and (\ref{induced_ee_offdiag})--(\ref{induced_te_offdiag}),
      for the various correlation functions.} 
\begin{ruledtabular}
\begin{tabular}{cc}
{\rm XX}$'$&$S_{l l'}^{L,{\rm X X'}}$\\\noalign{\smallskip}\hline\noalign{\smallskip}
TT&$ \left(C_{l'}^{\rm T,dT}+C_{l}^{\rm T,dT}\right)K_{ll'}^{L}$  \\\noalign{\smallskip}
EE&$ \left(C_{l}^{\rm E, dE}+C_{l'}^{\rm E, dE}\right)H^{L}_{l l'}$  \\ \noalign{\smallskip}
EB&$ -iC_{l'}^{\rm E, dE}H^{L}_{l l'} $   \\ \noalign{\smallskip}
TB&$-iC_{l'}^{\rm T, dE}H^{L}_{l l'} $\\\noalign{\smallskip}
TE&$ \left(C_{l'}^{\rm T,dE}H^{L}_{l l'}+C_{l}^{\rm E,dT}K^{L}_{l l'}\right)$
\end{tabular}
\end{ruledtabular}
\label{mfunc}
\end{table}

The spherical-harmonic coefficients $X_{lm}^{\mathrm{map}}$
obtained by a given CMB experiment are related to the true
coefficients $X_{lm}$ by $X_{lm}^{\rm map}=W_{l}X_{lm}$, where
$W_{l}=e^{-l\left(l+1\right)\sigma_{b}^{2}/2}$ is the window
function, and $\sigma_{b} =\theta_{\rm fwhm}/\sqrt{8\ln{2}}\simeq
0.00741 \left(\theta_{\rm fwhm}/1^{\circ}\right)$ is related to
the beam's full width at half maximum (FWHM) $\theta_{\rm
fwhm}$. The observed two-point correlations are then
\begin{align}
     \left \langle X^{\prime,{\rm
     map}*}_{l'm'} X_{lm}^{\rm
     map}\right\rangle=&\\
     C_{l}^{\rm X
     X'} W_{l}^{2}\delta_{l
     l'}\delta_{m m'}&+\sum_{LM}D^{LM,{\rm X
     X',map}}_{l
     l'}\xi^{LM}_{lm,l'm'},\nonumber\\
     D^{LM,{\rm X X',map}}_{l l'}=&~D^{LM,{\rm X
     X'}}_{l l'}W_{l}W_{l'}.
\end{align}
Following
Refs.~\cite{kamion_derotate,pullen_kamion,dvorkin_smith_a,dvorkin_smith_b,gluscevic_kamion_fullsky,gluscald},
the minimum-variance quadratic estimator for the rotational
invariant $D^{LM,{\rm X X'}}_{l l',{\rm map}}$ is
\begin{equation}
     \widehat{D}^{LM,{\rm X X'},{\rm map}}_{l
     l'}=\left(G_{l l'}\right)^{-1}\sum_{m
     m'}X_{lm}^{\rm
     map} X_{l'm'}^{\prime,{\rm
     map}*}
     \xi^{LM}_{lm,l'm'}.
\label{optimal_d}
\end{equation}
To extract $\Delta_{LM}$, CMB temperature and polarization maps must be
used to reconstruct $\widehat{D}^{LM,{\rm X X'},{\rm
map}}_{l l'}$ by applying Eq.~(\ref{optimal_d}). Then,
through the estimator $\widehat{\Delta}_{LM}^{l l',{\rm X
X'}}\equiv\widehat{D}^{LM,{\rm X X',map}}_{l
l'}/\left(W_{l}W_{l'}S_{l l'}^{L,{\rm X
X'}}\right)$, we obtain many measurements of
$\Delta_{LM}$. These measurements are generally correlated (even
for fixed $l,l')$, so we must take care to construct an
optimal estimator $\widehat{\Delta}_{LM}$ for CIPs when
using the full set of available maps for a given
experiment. Generalizing the estimator and error formulae in
Refs.~\cite{pullen_kamion,gluscevic_kamion_fullsky} to our case
of interest, we obtain the optimal estimator $\widehat{\Delta}_{LM}$
and its rms error $\sigma_{\Delta_{L}}$, taking into account
all possible correlations between X and X$'$: 
\begin{align}
     \widehat{\Delta}_{LM}=&~\sigma_{\Delta_{L}}^{2}\sum_{l'\geq
     l} G_{l l'}\sum _{\rm A A'} \mathcal{Z}_{l
     l'}^{L,{\rm A'}} \widehat{D}_{l
     l'}^{LM,{\rm A},{\rm map}}\left[\mathcal{C}_{l
     l'}^{-1}\right]_{\rm A A'},\nonumber\\ \sigma_{\Delta_{L}}^{-2}=&~\sum_{l'\geq l} G_{l
     l'} \sum _{\rm A A'} \mathcal{Z}_{l
     l'}^{L,{\rm A'}} \mathcal{Z}_{l l'}^{L,{\rm
     A}}\left[\mathcal{C}_{l l'}^{-1}\right]_{\rm A
     A'},
\label{toterr}
\end{align}
where $\left\{A, A'\right\}\in \left\{{\rm EB,BE,TB,BT,TT,EE,TE,ET
}\right\}$ when $l\neq l$, $\left\{A, A'\right\}\in \left\{{\rm
EB,TB,TT,EE,TE}\right\}$ when $l=l'$, and 
\begin{equation}
\mathcal{Z}_{l l'}^{L,{\rm A}}\equiv S_{l l'}^{L,{\rm A}}W_{l}W_{l'}.
\end{equation} The inequality
$l'\geq l$ is imposed so that we do not double count
correlations.  Sums are subject to the additional restriction that for $\left\{A, A'\right\}\in (\rm TE, ET, EE, TT)$, $l+l^{\prime}+L$ is even, while for $\left\{A, A'\right\}\in (\rm BE, EB, BT, TB)$, $l+l^{\prime}+L$ is odd. The appropriately normalized covariance matrix
for $\widehat{D}_{l l'}^{LM,{\rm A A'},{\rm
map}}$ is
\begin{align}
     \mathcal{C}_{l l'}^{\rm A A'}\equiv &~G_{l
     l'}\left(\left \langle \widehat{D}_{l
     l'}^{LM,{\rm A,map}}\widehat{D}_{l l'}^{LM,{\rm
     A',map*}}\right\rangle \right.\label{covar_mat}\\
     &-\left.\left\langle\widehat{D}_{l l'}^{LM,{\rm
     A,map}}\right\rangle\left\langle\widehat{D}_{l
     l'}^{LM,{\rm
     A',map*}}\right\rangle\right).\nonumber
\end{align}

We now proceed to compute the covariance
matrix $\mathcal{C}_{l l'}^{\rm A \rm A'}$.

\subsection{Off-diagonal covariances}
\label{Dcovar_sec}

To numerically evaluate Eq.~(\ref{covar_mat}), we must have a
model for the statistics of the observed map covariances,
\textit{including} noise. We assume the noise in each pixel is
statistically independent, Gaussian, and uncorrelated with the
signal, and we assume no coupling between the noises in
$\left\{{\rm T,~E,~B}\right\}$. In this case, the noise power
spectra are \cite{knox}
\begin{eqnarray}
     C_{l}^{\rm BB, noise}=C_{l}^{\rm EE, noise}&=&2 C_{l}^{\rm
     TT, noise}\nonumber \\&= &8\pi \frac{f_{\rm survey}({\rm NET})^{2}}{t_{\rm obs}},
\label{altnoise}
\end{eqnarray}
where NET is the (effective) noise-equivalent temperature for
the experiment, $t_{\rm obs}$ the duration of
the experiment, and $f_{\rm survey}$ is the fraction of sky
surveyed.  The cross spectra $C_{l}^{\rm X X',noise}=0$
if ${\rm X}\neq {\rm X}'$.  The power spectra for the
map are
\begin{equation}
     C^{XX'\text{, map}}_l\equiv C^{XX'}_l|W_l|^2 +
     C^{XX'\text{, noise}}_l.
\label{Cobs}
\end{equation}

It is useful to explicitly evaluate Eq.~(\ref{covar_mat}) using
Eq.~(\ref{optimal_d}).  Since all fields involved are Gaussian,
all the four-point functions that arise may be evaluated using
Wick's theorem. Wigner-$3J$ identities may then
be fruitfully applied to obtain all the elements of
$\mathcal{C}_{l l'}^{\rm A \rm A'}$ expressed in
terms of $C_{l}^{\rm X X',map}$. If $l=l'$, then
$\mathcal{C}_{ll}$ is a $5\times5$ diagonal matrix, with
rows/columns in the order TT, EE, TE, BE, BT, and entries
\begin{widetext}
\begin{align}
     \mathcal{C}_{l l}=&\left(\begin{array}{ll} \mathcal{F}_{l
     l}&\mathbf{0}\\ 
     \mathbf{0}^{*}&\mathcal G_{l
     l}\end{array}\right),~~\mathcal{F}_{l
     l}=2\left( \begin{array}{lll}\left(C_{l}^{\rm
     TT,map}\right)^{2}&\left(C_{l}^{\rm TE,map}\right)^{2}
     &C_{l}^{\rm TT,map} C_{l}^{\rm TE,map}\\ \left(C_{l}^{\rm
     TE,map}\right)^{2}&\left(C_{l}^{\rm EE,map}\right)^{2}
     &C_{l}^{\rm EE,map} C_{l}^{\rm TE,map} \\
     C_{l}^{\rm TT,map} C_{l}^{\rm TE,map}&C_{l}^{\rm EE,map}
     C_{l}^{\rm TE,map}& \left[\left(C_{l}^{\rm
     TE,map}\right)^{2}+C_{l}^{\rm TT,map}C_{l}^{\rm
     EE,map}\right]/2\end{array}\right),
\end{align}
\end{widetext}
\begin{align}
     \mathcal{G}_{l l}=&\left(\begin{array}{ll}C_{l}^{\rm
     EE,map}C_{l}^{\rm BB,map}&C_{l}^{\rm BB,map}C_{l}^{\rm
     TE,map}\\\\C_{l}^{\rm BB,map}C_{l}^{\rm TE,map}&C_{l}^{\rm
     BB,map}C_{l}^{\rm TT,map}\end{array}\right).
\end{align}
If $l\neq l'$, $\mathcal{C}_{l l'}$ is an
$8\times8$ block-diagonal matrix, with rows/columns in the order
TT, EE, TE, ET, BE, EB, BT, TB, and entries
\begin{widetext}
\begin{align}
     \mathcal{C}_{l
     l'}=\left(\begin{array}{ll}\mathcal{N}_{l
     l'}&\mathbf{0}\\ \mathbf{0}&\mathcal{K}_{l
     l'} \end{array}\right),~~~
     \mathcal{N}_{l
     l'}=\left(\begin{array}{llll}C_{l}^{\rm
     TT,map}C_{l'}^{\rm TT,map}&C_{l}^{\rm
     TE,map}C_{l'}^{\rm TE,map}&C_{l}^{\rm
     TT,map}C_{l'}^{\rm TE,map}&C_{l}^{\rm
     TE,map}C_{l'}^{\rm TT,map}\\ \\
     C_{l}^{\rm TE,map}C_{l'}^{\rm TE,map}&C_{l}^{\rm
     EE,map}C_{l'}^{\rm EE,map}&C_{l}^{\rm
     TE,map}C_{l'}^{\rm EE,map}&C_{l}^{\rm
     EE,map}C_{l'}^{\rm TE,map}\\ \\
     C_{l}^{\rm TT,map}C_{l'}^{\rm TE,map}&C_{l}^{\rm
     TE,map}C_{l'}^{\rm EE,map}&C_{l}^{\rm
     TT,map}C_{l'}^{\rm EE,map}&C_{l}^{\rm
     TE,map}C_{l'}^{\rm TE,map}\\ \\
     C_{l}^{\rm TE,map}C_{l'}^{\rm TT,map}&C_{l}^{\rm
     EE,map}C_{l'}^{\rm TE,map}&C_{l}^{\rm
     TE,map}C_{l'}^{\rm TE,map}&C_{l}^{\rm
     EE,map}C_{l'}^{\rm TT,map}
      \end{array}\right),\\\nonumber\\
       \mathcal{K}_{l
       l'}=\left(\begin{array}{cccc}C_{l}^{\rm
       BB,map}C_{l'}^{\rm EE}&0&C_{l}^{\rm
       BB,map}C_{l'}^{\rm TE,map}&0\\ \\  0&C_{l}^{\rm
       EE,map}C_{l'}^{\rm BB,map}&0&C_{l}^{\rm
       TE,map}C_{l'}^{\rm BB,map}   \\ \\   C_{l}^{\rm
       BB,map}C_{l'}^{\rm TE,map}&0&C_{l}^{\rm
       BB,map}C_{l'}^{\rm TT,map}&0   \\ \\ 
       0&C_{l}^{\rm TE,map}C_{l'}^{\rm
       BB,map}&0&C_{l}^{\rm TT}C_{l'}^{\rm BB,map}
\end{array}\right)\nonumber.
\end{align}
\end{widetext}
In Sec.~\ref{noisecurves}, we apply the preceding formulae to
estimate the noise in the reconstructed CIP field for a variety
of ongoing and upcoming experiments. 

In many cases, most of the sensivity to CIPs comes from a single
combination (e.g., TT or TB) of observables.  It is therefore
interesting to consider the constraining power of a single such
combination.  In the case of TB, the error is given by [see Eqs.~(\ref{toterr})]
\begin{equation}
     \sigma_{\Delta_{L}}^{-2}=\sum_{l'\geq l}^{l+l^{\prime}+L~{\rm odd}} G_{l
     l'} \frac{\left(S_{l l'}^{L,{\rm TB}}\right)^{2}}{C_{l}^{\rm
     BB,map}C_{l'}^{\rm TT,map}}+~~\left\{{\rm T
     \leftrightarrow B}\right\}.
\label{singleatatime}
\end{equation}
To generate the noise curves discussed in
Sec.~\ref{noisecurves}, we use expressions analogous to
Eq.~(\ref{singleatatime}) for each pair of observables. These
noise values are then applied to estimate the SNR with which a given spectrum of CIPs
might be detected.

\subsection{Signal-to-noise formula}
\label{SN_sec}

Ultimately, we wish to assess the SNR of our estimators for a
given CIP power-spectrum amplitude $A$.  Each estimator
$\widehat\Delta_{LM}$ gives an independent estimator for
$A$, and by adding them all with inverse-variance weighting, the
total SNR with which CIPs can be detected is
\begin{equation}
     S/N=\left[\frac{f_{\rm
     sky}}{2}\sum_{L>L_{\rm min}}\left(2L+1\right)
     \left(\frac{C_{L}}{\sigma_{\Delta_{L}}^{2}} \right)^{2}
     \right]^{1/2},
\label{snratexp}
\end{equation} 
where the error is evaluated using Eq.~(\ref{toterr}), $f_{\rm sky}$ is the fraction of sky used in the data analysis, and $L_{\rm min}\equiv f_{\rm sky}^{-1/2}$. Modes that vary on scales larger than the area of sky analyzed will have degraded signal-to-noise. A minimum value of $L$ is thus imposed to conservatively account for fractional sky coverage. In practice, these modes would still contribute to the integrated CIP power in the area of sky analyzed. In this work, however, we impose a cut at $L_{\rm min}$ to avoid an over-estimate of sensitivity, all the same establishing the utility of CMB observations for probing CIPs.

\section{Experimental prospects}
\label{exprop}

We now apply the formalism of Sec.~\ref{qestsecbig} to assess
the prospects of using CMB experiments to detect CIPs. We consider specifically the ongoing WMAP
\cite{wmap1_komatsu} and Planck \cite{planck_bluebook}
satellites and a possible future satellite, EPIC \cite{cmbpol}.
We also consider the following suborbital experiments: 
Polarbear \cite{polarbear}, Spider \cite{spider} and ACT
\cite{act} and SPT \cite{spt} and their polarization upgrades
ACTPol \cite{actpol} and SPTPol \cite{sptpol}.  Finally we consider an idealistic cosmic-variance limited (CVL) experiment, limited only by sky cuts to avoid galactic foreground emission. We do this to roughly quantify the lowest CIP amplitudes that could ever be probed with the CMB. 

The experimental parameters assumed for these experiments are given in 
Table~II. For WMAP, Planck, EPIC, and the CVL case, we assume that $f_{\rm survey}=1$, while for Polarbear, SPT, ACT, SPTPol, and ACTPol, we assume that $f_{\rm survey}=f_{\rm sky}$. For WMAP, we assume use of the V and W bands in the analysis. For Planck, we assume that the $143$ and $217$ Ghz channels will be usable for CMB anisotropy measurements, and take appropriate inverse-variance weighted sums of the noise in these channels. Appendix \ref{deriv_pspec_sec} details the calculation of the requisite derivative power spectra. We include BB correlations induced by gravitational lensing when evaluating $\mathcal{C}_{l l^{\prime}}^{\rm A A^{\prime}}$, using the \textsc{camb}  lensing module \cite{camb}.

\begin{table}[htbp]
\begin{ruledtabular}
\begin{center}
\begin{tabular}{cccccc} \hline \\
Expt.&Channel&$\theta$&NET&$f_{\rm sky}$&$t_{\rm obs}$\\ \hline 
WMAP&V Band&$21$&$1200$&$0.65$&$7$\\ \hline
WMAP&W Band&$13$&$1600$&$0.65$&$7$\\ \hline
Polarbear&$150~{\rm Ghz}$&$4.0$&$36$&$0.015$&$1.0$\\ \hline 
Planck HFI&$143~{\rm Ghz}$&$7.1$&$62$&$0.65$&$1.2$\\ \hline
Planck HFI&$217~{\rm Ghz}$&$5.0$&$91$&$0.65$&$1.2$\\ \hline 
Spider&$150~{\rm Ghz}$&$30$&$4.2$&$0.1$&$0.016$\\ \hline 
ACT&$148~{\rm Ghz}$&$1.4$&$58$&$0.0072$&$0.14$\\ \hline
SPT&$150~{\rm Ghz}$&$1.2$&$91$&$0.0024$&$0.29$\\ \hline 
ACTPol&$150~{\rm Ghz}$&$1.4$&$6.0$&$0.10$&$0.21$\\ \hline 
SPTPol&$150~{\rm Ghz}$&$1.0$&$14$&$0.016$&$0.75$\\ \hline 
EPIC&$150~{\rm Ghz}$&$5.0$&$2.0$&$0.65$&$4.0$\\  \hline
CVL&$...$&$...$&$0.0$&$0.65$&$...$\\  
\hline
\end{tabular}\end{center}
\caption{Experimental parameters for the experiments considered in
     this work: beamwidth $\theta$ (in arcminutes),
     noise-equivalent temperature (NET) (in $\mu$K~sec$^{1/2}$),
     and observation time $t_{\text{obs}}$ (in years).}
\end{ruledtabular}
\label{exp_parameters}
\end{table}


\begin{figure*}[htbp]
\includegraphics[width=7.0in]{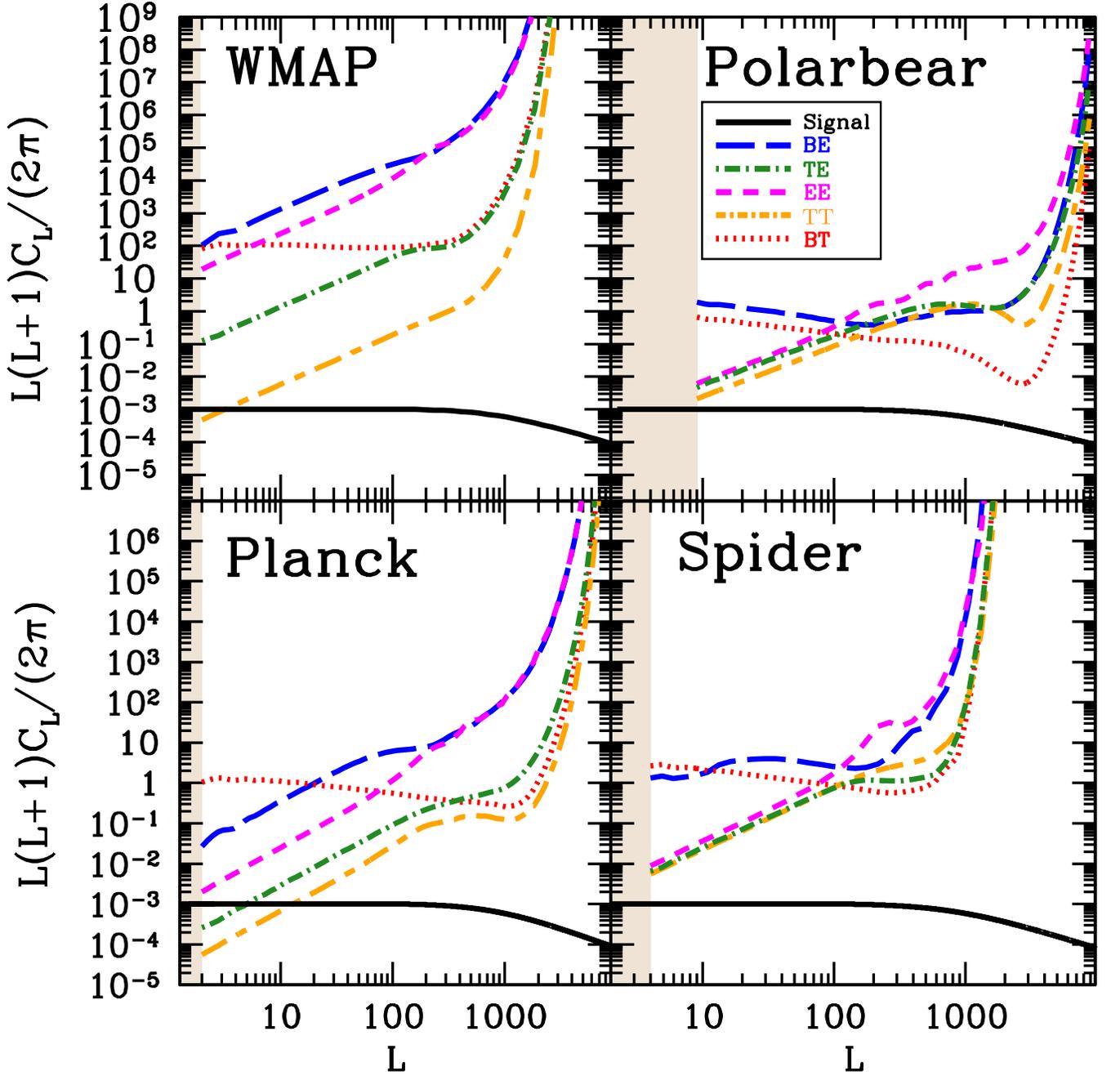}
\caption{Predicted noise power spectrum $\delta C_{L}$ in the reconstructed CIP
     perturbation field $\Delta_{LM}$ in four different
     ongoing/future CMB anisotropy experiments, as a function of
     angular scale $L$. We separately plot the noise for
     distinct pairs of observables: TT is shown as an orange (short-dashed long-dashed), TE as a green (dotted-dashed) line, EE as a
     magenta (short-dashed) line, BE as a blue (long-dashed)
     line, and BT as a red (dotted) line. Also shown (black solid line) is the power spectrum
     $C_L^\Delta$ , marked signal, for a scale-invariant spectrum of CIPs with
     the maximum amplitude allowed by galaxy clusters. Each panel shows
     estimates for a different experiment, as indicated in the
     figure. The beige (grey) shaded region shows the range of $L$ that is not included in our estimates, due to finite sky coverage effects, as discussed in Sec. \ref{SN_sec}.}
\label{noise_a}
\end{figure*}

\begin{figure*}[htbp]
\includegraphics[width=7.0in]{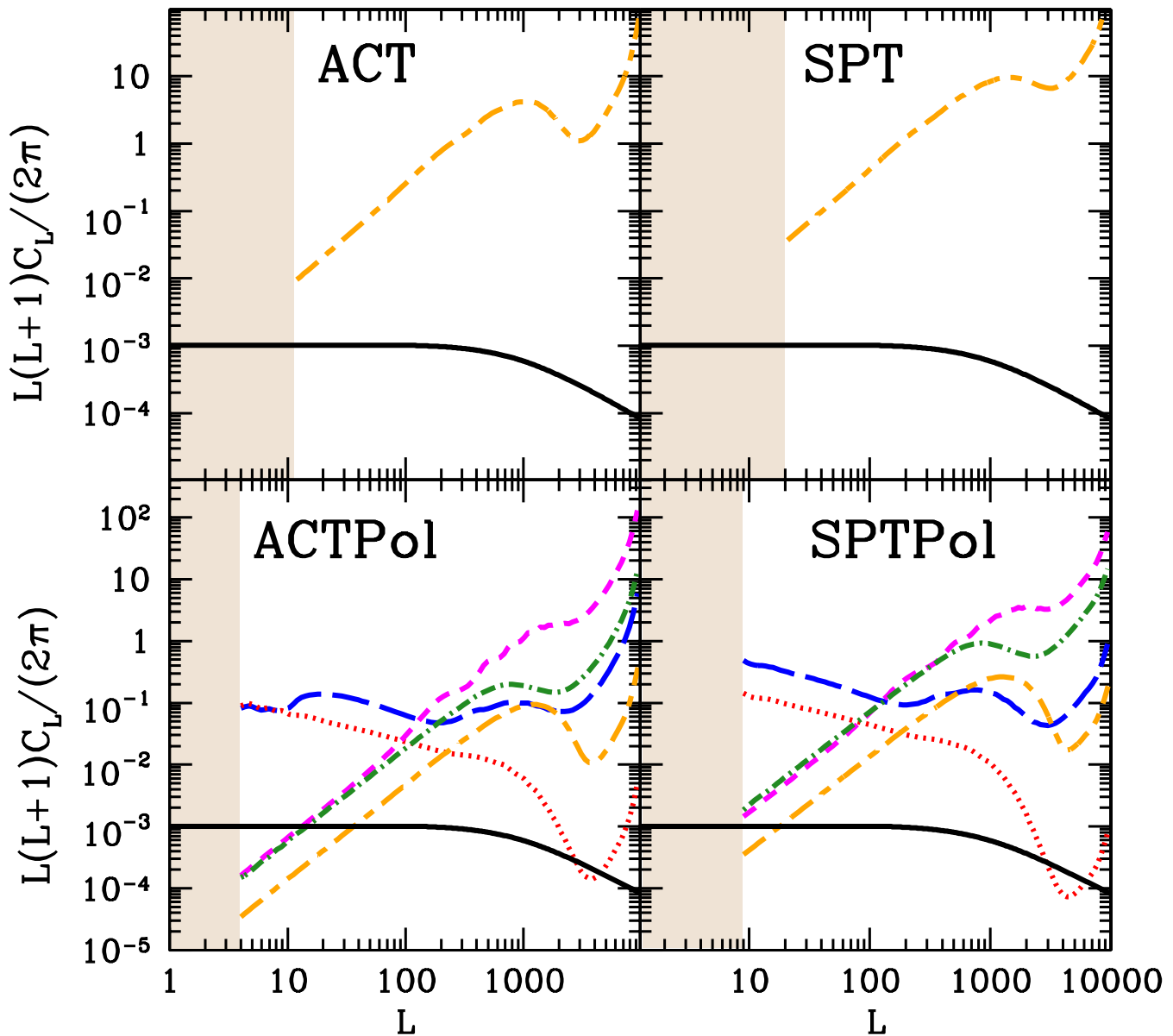}
\caption{Predicted noise power spectrum $\delta C_{L}$ in the reconstructed CIP
     perturbation field $\Delta_{LM}$ in four different
     ongoing/future CMB anisotropy experiments, as a function of
     angular scale $L$. Colors (line styles) are as in Fig. \ref{noise_a}. We separately plot the noise for
     distinct pairs of observables: TT is shown as an orange (short-dashed long-dashed), TE as a green (dotted-dashed) line, EE as a
     magenta (short-dashed) line, BE as a blue (long-dashed)
     line, and BT as a red (dotted) line. Also shown (black solid line) is the power spectrum
     $C_L^\Delta$ for a scale-invariant spectrum of CIPs with
     the maximum amplitude allowed by galaxy clusters. Each panel shows
     estimates for a different experiment, as indicated in the
     figure. The beige (grey) shaded region shows the range of $L$ that is not included in our estimates, due to finite sky coverage effects, as discussed in Sec. \ref{SN_sec}.}
\label{noise_b}

\end{figure*}
\begin{figure}[htbp]
\includegraphics[width=3.26in]{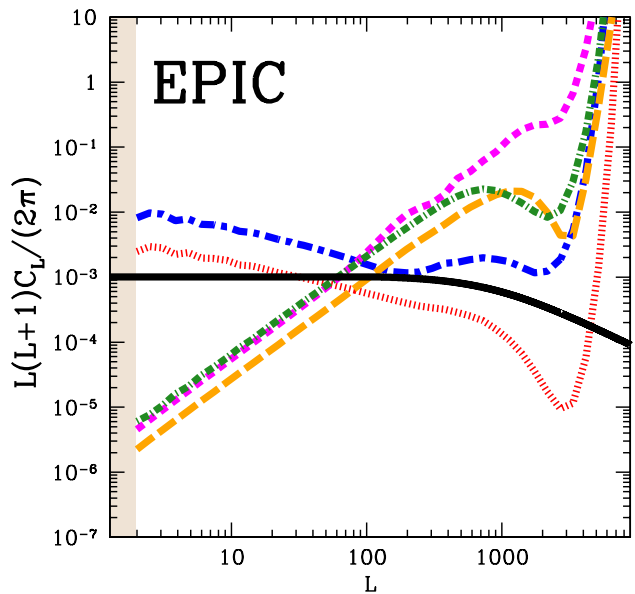}
\caption{Predicted noise power spectrum $\delta C_{L}$ in the reconstructed CIP
     perturbation field $\Delta_{LM}$ for the proposed EPIC
     satellite, as a function of angular scale $L$. Colors (line styles) are as in Fig. \ref{noise_a}. We separately plot the noise for
     distinct pairs of observables: TT is shown as an orange (short-dashed long-dashed), TE as a green (dotted-dashed) line, EE as a
     magenta (short-dashed) line, BE as a blue (long-dashed)
     line, and BT as a red (dotted) line. Also shown (black solid line) is the power spectrum
     $C_L^\Delta$ for a scale-invariant spectrum of CIPs with
     the maximum amplitude allowed by galaxy clusters. The beige (grey) shaded region shows the range of $L$ that is not included in our estimates, due to finite sky coverage effects, as discussed in Sec. \ref{SN_sec}.}
\label{noise_c}
\end{figure}

\begin{figure}[htbp]
\includegraphics[width=3.35in]{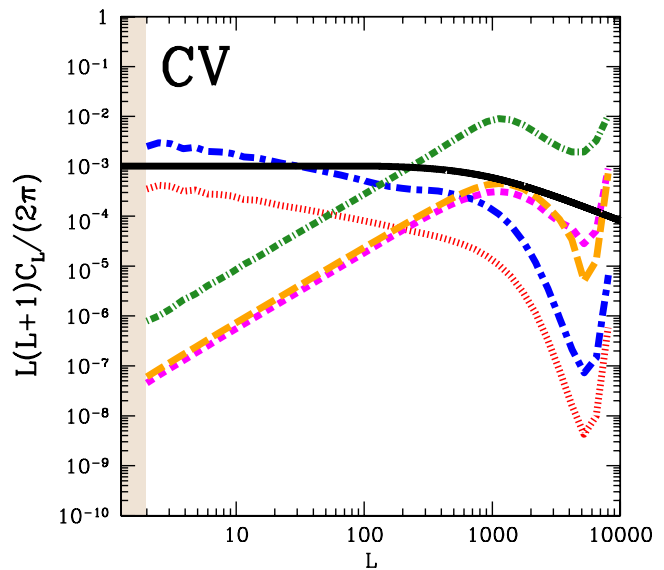}
\caption{Predicted noise power spectrum $\delta C_{L}$ in the reconstructed CIP
     perturbation field $\Delta_{LM}$ for an ideal cosmic variance limited experiment, as a function of angular scale $L$. Colors (line styles) are as in Fig. \ref{noise_a}. We separately plot the noise for
     distinct pairs of observables: TT is shown as an orange (short-dashed long-dashed), TE as a green (dotted-dashed) line, EE as a
     magenta (short-dashed) line, BE as a blue (long-dashed)
     line, and BT as a red (dotted) line. Also shown (black solid line) is the power spectrum
     $C_L^\Delta$, for a scale-invariant spectrum of CIPs with
     the maximum amplitude allowed by galaxy clusters. The beige (grey) shaded region shows the range of $L$ that is not included in our estimates, due to finite sky coverage effects, as discussed in Sec. \ref{SN_sec}.}
\label{noise_d}
\end{figure}

\begin{figure*}[htbp]
\centering
 \includegraphics[width=6.900in]{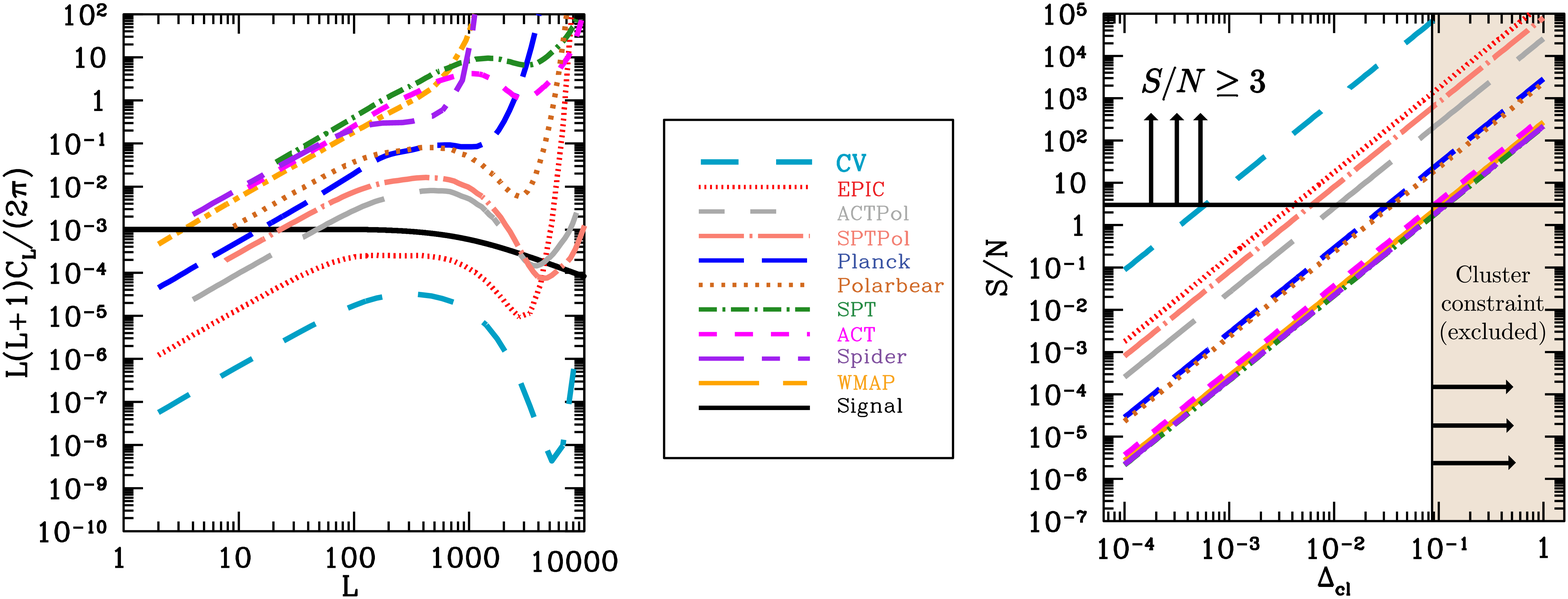}
\caption{Combined (predicted) noise power spectrum $\delta C_{L}$ in the reconstructed CIP perturbation field $\Delta_{LM}$ for $9$ different CMB experiments, as a function of angular scale $L$. Colors and line styles for each experiment are indicated in the legend (center). Here, noise for the $5$ different estimators (TT, TE, EE, BE, and BT) is added in quadrature. Noise curves terminate at L values where modes become inaccessible due to finite sky effects, as discussed in Sec. \ref{SN_sec}.
Also shown (black solid line) is the power spectrum
     $C_L^\Delta$ , marked signal, for a scale-invariant spectrum of CIPs with
     the maximum amplitude allowed by galaxy clusters. The left panel shows predicted total
     errors for the indicated 
     experiments. 
     The right panel shows the predicted signal-to-noise ratio
     (SNR) that results from these errors, evaluated using
     Eq.~(\ref{snratexp}) and assuming a scale-invariant
     spectrum of CIPs [evaluated using
     Eq.~(\ref{precise_cL})]. The SNR is plotted as a function
     of the rms CIP fluctuation $\Delta_{\rm cl}$ on cluster scales. The range of fluctuations
     $\Delta_{\rm cl}$ excluded by cluster
     measurements~\cite{holder} is shown as a beige (grey) band, bounded by a vertical black line with rightward pointing arrows. The black line with upward arrows attached shows the ``detection" region, defined by $S/N=3$.}
\label{moneyplot}
\end{figure*}

\subsection{Noise curves}
\label{noisecurves}

We compute the noise curves in the reconstruction
$\widehat{\Delta}_{LM}$ for all experiments under consideration, and show the results in Figs.~\ref{noise_a}--\ref{noise_d}. We plot the noise power spectrum\begin{equation}
\delta C_{L}=\frac{\sigma_{\Delta_{L}}^{2}}{f_{\rm sky}\sqrt{2L+1}}\end{equation} as well as the signal $C_{L}$. We use the value $A=0.017$, which saturates the galaxy-cluster bound on CIPs. Experiments with
larger beams, such as Polarbear and Spider, generally have
higher noise levels (for reconstruction of $\Delta$) than do
others. At low $L$, temperature is generally the best probe of
CIPs. At sufficiently high $L$, the BT correlation takes
over as the best probe of CIPs. At very high $L$,
all the noise curves grow very large, indicating that the
fidelity of the reconstruction of the CIP
breaks down at small scales. This is very similar to
lensing and is expected, as fluctuations in the baryon fraction
couple small to large scales. We note that $C_{l}^{{\rm X} {\rm X^{\prime}},{\rm map}}$ is computed using ${\tau}=0.086$, as the observed CMB anisotropies are affected both by a screening envelope at high $l$, due to reionization, and by a reionization ``bump" in polarization at low $l$. We use \textsc{camb}'s built-in \textit{tanh} reionization model with the parameters of Ref. \cite{lewis_reion}.

We then compute the total noise in
the reconstruction $\widehat{\Delta}_{LM}$ for each experiment,
adding the different correlations in quadrature with inverse-variance weighting. This should be a reasonable approximation to
the sum in Eq.~(\ref{toterr}), as inverse-variance weighting
tends to be dominated sharply by the lowest noise
correlation. The results are shown in the left panel of
Fig.~\ref{moneyplot}. We see that broadly speaking (with some
alternation as a function of $L$), the best sensitivity is
achieved by EPIC, followed by SPTPol, ACTPol, Planck, Polarbear,
WMAP, ACT, Spider, and SPT. Obviously, any specific experimental concept will be outperformed by the CVL case, as confirmed in the left panel of Fig. \ref{moneyplot}.

\subsection{Signal-to-noise}
\label{snres}
Calling on Eq.~(\ref{snratexp}), we compute the SNR expected for
all the CMB experiments we consider as a
function of the rms CIP fluctuation $\Delta_{\rm cl}$. The
results are shown in the right panel of
Fig.~\ref{moneyplot}.  Assuming a scale-invariant spectrum, we
see that already with WMAP, we are able to probe compensated
fluctuations in the baryon fraction of $\sim 10\%$, the lowest
value probed in Ref.~\cite{holder}. We define a ``detection" as a
measurement with $S/N\geq 3$. Currently operating suborbital experiments, like 
SPT and ACT, and the upcoming Spider experiment, perform comparably to WMAP. Although these experiments are sensitive to a fairly large rms CIP amplitude, the cluster constraint was obtained at a different scale, and it is important to check the constraint using the independent probe offered by the CMB.

Planck and Polarbear offer the next
major improvement in SNR, probing $\Delta_{\rm cl}\simeq
3\%$ and higher. The addition of polarization sensitivity to SPT and ACT
lowers the range of detectable $\Delta_{\rm cl}$ to
$0.6\%$. EPIC would be able to measure $\Delta_{\rm
cl}\simeq0.4\%$. This is a factor of $20$ lower than the currently allowed
maximum from measurements of the baryon fraction in galaxy clusters. In the CVL case, an additional order of magnitude improvement in SNR is possible. 

As discussed in Sec. \ref{SN_sec}, we conservatively estimated sensitivity, omitting low-L modes. In practice, all-sky experiments like WMAP, Planck, and EPIC could probe even smaller CIP amplitudes than estimated here, as the $L=1$ mode could contribute significantly to the SNR. If experimental techniques improve (approaching the cosmic-variance limit at very high $l$) and a way is found to disentangle CIPs from secondary CMB anisotropies at these high $l$, a further order of magnitude improvement in sensitivity is theoretically possible, using the estimators developed in this work. Additionally, we have been conservative in our estimates of SNR, assuming that only one useful frequency channel is available for EPIC. The mission concept actually calls for $\sim 7$ channels, in order to achieve good foreground subtraction. It may be that if EPIC is built, more than one useful channel of signal is obtained, improving the SNR by a factor of order unity.

It may be that CIPs are not an independent Gaussian random field (as assumed here), but rather, as in some curvaton models \cite{gordon_pritchard}, correlated with the usual adiabatic fluctuations. In that case, CIPs will induce $3$-point correlations between CMB observables, through effects at decoupling. In future work, we will pursue the possibility of probing CIPs with the corresponding CMB bispectra.

\section{Conclusions}
\label{conc}

Compensated isocurvature perturbations provide an intriguing
empirical possibility for large-amplitude departures from
homogeneity in the early Universe.  The current constraint,
$\lesssim 10\%$, to the amplitude of such perturbations is
surprisingly weak.  While Ref.~\cite{holder} has pointed out
that there may be CMB signatures induced by CIPs at reionization,
we have shown here that the CMB effects of CIPs at the surface of
last scatter would be several orders of magnitude larger.  We then calculated the
full two-point temperature/polarization correlations induced by
CIPs on the CMB and developed the minimum-variance estimators
for measuring the CIP field with the CMB.

The WMAP satellite may be sensitive to a scale-invariant
spectrum of CIPs, but only if the CIP amplitude is close to its
current upper bound. In the future, sensitivity to CIP amplitudes as small as $\sim 3 \%$ may be achieved by instruments in operation and $\sim
0.1\%$-level fluctuations accessible in the near future with
precise ground- and space-based polarization experiments that are under construction (ACTPol and SPTPol) or in conceptual development (EPIC). 

Many steps must be taken before such measurements can be
implemented with real data.  Techniques to deal with partial-sky
coverage and realistic instrumental noise properties must be
developed, but these techniques should be similar to those
being developed already to measure the effects of weak
gravitational lensing on the CMB.  Likewise, techniques must be
developed to distinguish the off-diagonal correlations induced
by CIPs from those induced by weak gravitational lensing (e.g., \cite{meng}), which should be comparable in amplitude if the CIP field $\Delta$ is comparable to the lensing potential $\phi$, i.e, $\sim 1\%$.

Although lensing is already included in our reconstruction noise estimates [$\sigma_{\Delta_{L}}$ in Eq. (\ref{toterr})], it might also induce bias in the reconstruction of the CIPs. Gravitational lensing of the CMB will induce correlations of similar form to Eqs.~(\ref{correlaa}) and (\ref{correlab}), with the distinction that the functions $S_{l l^{\prime}}^{L,{\rm X X^{\prime}}}$ will be different for lensing than for CIPs. In the case of lensing, these functions describe the remapping of CMB observables on a lensing-deflected sky. In the case of CIPs, these functions describe the detailed physical dependence of CMB anisotropies on the baryon density. If Eq.~(\ref{toterr}) is applied to the extension of Eqs.~(\ref{correlaa})-(\ref{correlab}) that includes gravitational lensing, a bias will be induced in the measurement of $\Delta_{LM}$.

The differing forms of $S_{l l^{\prime}}^{L,{\rm X X^{\prime}}}$ will allow lensing and CIPs to be disentangled. Using a straightforward generalization of the estimator in Eq.~(\ref{toterr}) that includes terms for both CIPs and lensing, the bias induced by lensing on CIP measurements may be removed, 
simultaneously reconstructing the lensing and CIP fields. This is analogous to the estimators discussed in Ref. \cite{meng}, where it is shown that if reionization is patchy because of inhomogeneity in the distribution of ionized bubbles around the first sources, the contributions of patchy reionization and lensing may be distinguished. In future work, we will explicitly compute the bias in CIP searches that will be induced by lensing and will construct the estimators that disentangle lensing from CIPs. As in Ref. \cite{meng}, we expect that the signal-to-noise of the biased and unbiased estimators should be nearly the same, and so gravitational lensing should not affect the signal-to-noise estimates of this paper.

The measurements we propose in this paper offer a precise test of how closely the primordial baryon and dark-matter distributions are matched and approach the CIP amplitudes allowed in curvaton models. Moreover, if future CMB experiments detect subdominant isocurvature fluctuations between matter and radiation, the techniques developed in this work could disentangle contributions from the baryon and cold dark-matter (CDM) isocurvature modes. Even greater gains in sensitivity are theoretically possible if the cosmic-variance limit is approached at high $l$ by future experiments. We are optimistic that, in the near future, we will learn just how well baryons trace dark matter in the early Universe.

\begin{acknowledgments}

We acknowledge useful conversations with C.~Chiang, 
C.~Dvorkin, G.~Holder, M.~LoVerde, K.~M.~Smith, T.~L.~Smith,
D.~N.~Spergel, and M.~Zaldarriaga. We thank
B.~Jones and A. Fraisse for providing updated parameters for
Spider forecasting. DG was supported at the Institute for
Advanced Study by the National Science Foundation
(AST-0807044) and is grateful for the hospitality of the Aspen Center for Physics, where part of this work was completed. MK thanks the support of the Miller Institute for Basic Research in Science and the hospitality of the
Department of Physics at the University of California, where
part of this work was completed.  This work was supported at
Caltech by DoE DE-FG03-92-ER40701, NASA NNX10AD04G, and the
Gordon and Betty Moore Foundation. Part of the research
described in this paper was carried out at the Jet Propulsion
Laboratory, California Institute of Technology, under a contract
with the National Aeronautics and Space Administration.

\end{acknowledgments}

  \setcounter{equation}{0} 
\appendix

\section{Tensors on the sphere}
 \renewcommand{\theequation}{A\arabic{equation}}
  \setcounter{equation}{0} 
  \label{tensor_on_sphere}
Here we review the tensor spherical-harmonic formalism,
following closely Ref.~\cite{gluscevic_kamion_fullsky}. The
metric on the 2-sphere is
\begin{equation}
     g=\tenspher{1}{0}{0}{\sin^{2}\theta},
\end{equation} 
where $\theta$ is the polar angle defined with respect to the
origin of the orientation vector $\hat{n}$. It is useful to
introduce the orthonormal basis
\begin{equation}\begin{array}{ll}
     \hat{e}_{\theta} = \left(\begin{array}{c} 1\\ 0 \end{array}
     \right) ~~~ \hat{e}_{\phi} = \left(\begin{array}{c} 0 \\\
     {\rm sin}^{2}\theta \end{array}\right).\end{array}
\end{equation} 
The tensor spherical harmonics are
\cite{kamion_polarization}
\begin{eqnarray}
     Y_{lm,ab}^{\rm E} = \frac{N_{l}}{2} \left( Y_{lm:ab} -
     \frac{1}{2} g_{ab} Y_{lm:c}^{c} \epsilon^{c}_{a} \right),\\
     Y_{lm,ab}^{\rm B} = \frac{N_{l}}{2} \left(Y_{lm:ac}
     \epsilon^{c}_{b} + Y_{lm:bc}\epsilon^{c}_{a}\right),
\end{eqnarray}
where the normalization constant is given by
\begin{equation}
     N_{l}\equiv\sqrt{\frac{2\left(l-2\right)!}{\left(l+2\right)!}},
\end{equation}
and all indices following ``$:$'' denote covariant derivatives
taken on the 2-sphere. The indices $l$ and $m$ are multipole
indices, while $a$ and $b$ are tensor indices.

Using Ref.~\cite{wayne_harmonic}, the covariant derivatives may
be expressed in terms of spin-2 spherical harmonics
\cite{wayne_harmonic,matias_uros_polarization}:
\begin{eqnarray}
     &&Y_{lm:ab} = -\frac{l\left(l+1\right)}{2} Y_{lm} g_{ab} +
     \frac{1}{2} \sqrt{\frac{ \left(l+2\right) }{ \left(l - 2
     \right)!}} \nonumber\\
     &&\times\left[\lsub{Y_{lm}}{2}\left(m_{+}\otimes
     m_{+}\right) + \lsub{ Y_{lm}}{-2} \left(m_{-}\otimes
     m_{-}\right)\right]_{ab}. 
\end{eqnarray}
The left subscript ``$2$'' denotes a spin-weighted spherical
harmonic $\lsub{Y_{lm}}{s}$ of spin $s$, while $\otimes$ denotes
a tensor product.  The spherical basis vectors  $m_{\pm}$ are
\begin{equation}
     m_{\pm} = \frac{1}{\sqrt{2}} \left(\hat{e}_{\theta}\mp i
     \hat{e}_{\phi}\right).
\end{equation}
In row-column form, the spherical tensor basis functions are
then \cite{gluscevic_kamion_fullsky}
\begin{align}
     Y^{\rm E} = \frac{ \tenspher{ \left(\lsub{Y}{+2} +
     \lsub{Y}{-2} \right)}{ i \sin{\theta} \left( \lsub{Y}{-2} -
     \lsub{Y}{+2} \right)} { i \sin{\theta} \left( \lsub{Y}{-2}
     - \lsub{Y}{+2} \right)}{ - \sin^{2}\theta
     \left(\lsub{Y}{-2} + \lsub{Y}{+2} \right)}}{2\sqrt{2}}, 
     \label{pbe}  \\ \nonumber \\
     Y^{\rm B} = \frac{\tenspher{i\left(\lsub{Y}{+2} -
     \lsub{Y}{-2}\right)} { \sin{\theta} \left(\lsub{Y}{-2} +
     \lsub{Y}{+2}\right)} { \sin{\theta} \left(\lsub{Y}{-2} +
     \lsub{Y}{+2} \right)}{ i \sin^{2}{\theta}
     \left(\lsub{Y}{-2} - \lsub{Y}{+2}\right)}}{2\sqrt{2}},
\label{pbb} 
\end{align} 
where we have suppressed the $lm$ indices for the sake of brevity.

To evaluate the polarization anistropies induced by CIPs (see
Sec.~\ref{polaniso}), it is useful to obtain identities for the
product of two distinct (generally different $l$ and $m$ values)
spherical harmonics. Using Eqs.~(\ref{pbe})--(\ref{pbb}), it can
be shown that \cite{gluscevic_kamion_fullsky}
\begin{eqnarray}
      \left(X^{{\rm E},ab} \right)^* Y_{ab}^{\rm E}&&\nonumber
      \\=\frac{1}{2}&\left(\lsub{X^{*}
      }{+2}\otimes\lsub{Y}{+2}+\lsub{X^{*}}{-2}\otimes
      \lsub{Y}{-2}\right),&
        \label{epair}\\
         \left(X^{{\rm B},ab} \right)^* Y_{ab}^{\rm E}&&\nonumber
         \\=-\frac{i}{2}&\left(\lsub{X^{*}
         }{+2}\otimes\lsub{Y}{+2}-\lsub{X^{*}}{-2}\otimes
         \lsub{Y}{-2}\right),&
\label{ebpair}
\end{eqnarray}
where $X$ and $Y$ signify tensor spherical harmonics on the left-hand side of Eqs.~(\ref{epair}) and (\ref{ebpair}) and the
corresponding spin-weighted spherical harmonics on the right-
hand side of these equations. Relations such as these help
express the integrals of Sec.~\ref{polaniso} as integrals over
$3$ spin-weighted spherical harmonics. These are then evaluated
by applying the relation \cite{wayne_harmonic}
\begin{eqnarray}
     &\int& d{\hat n}
     \left(\lsub{Y_{l_{1}m_{1}}^{*}}{ s_{1}}\right)
     \left(\lsub{Y_{l_{2}m_{2}}} {s_{2}} \right)
     \left(\lsub{Y_{l_{3}m_{3}}} {s_{3}}\right)\nonumber\\
     &=&\left(-1\right)^{m_{1}+s_{1}} \sqrt{\frac{\left(2l_{1} +
     1\right)\left(2l_{2} + 1\right) \left(2l_{3} + 1\right)}
     {4\pi}}\nonumber\\
     &\times&
     \wigner{l_{1}}{s_{1}}{l_{2}}{-s_{2}}{l_{3}}{-s_{3}}
     \wigner{l_{1}}{-m_{1}}{l_{2}}{m_{2}}{l_{3}}{m_{3}}.
\label{spinint}
\end{eqnarray}

In evaluating the contribution of terms proportional to
$d^{2}f^{\left(0\right)}/d\Delta^{2}$ to perturbed LOS solutions
for polarization anisotropies, integrals over 4 spin-weighted
spherical harmonics must be evaluated. They may be simplified
using the identity \cite{totalang}
\begin{eqnarray}\nonumber
     \lsub{Y_{l''
     -m''}}{-s}  \otimes \lsub{Y_{l'
     m'}}{s} = \sqrt{\left(2l'' + 1\right)
     \left( 2l' + 1\right)} \nonumber\\
     \times \sum_{L''M'' S''}
     \left(-1\right)^{M''+S''}
     \sqrt{\frac{2L''+1}{4\pi}}
     \lsub{Y_{L''M''}}{S''}\nonumber\\
      \times \wigner{l''}{m''}
     {l'}{-m'}{L''}
     {M''}
     \wigner{l''}{-s}{l'}{s}
     {L''}{-S''}.
\label{4spin}  
\end{eqnarray}

\section{Derivative power spectra}
 \renewcommand{\theequation}{B\arabic{equation}}

  \setcounter{equation}{0} 
\label{deriv_pspec_sec}

We wish to estimate the derivatives
$d^{n}X_{l}\left(k\right)/d\Delta^{n}$ of the transfer functions
$X_{l}\left(k\right)$.  For the first derivatives, we use a 5-point
approximation, running \textsc{camb} \cite{camb} repeatedly to
obtain the transfer functions at $5$ different values of
$\Omega_bh^{2}$ and $\Omega_ch^{2}$. The first
derivative is then well approximated by 
\begin{equation}
\frac{dX_{l}\left(k\right)}{d\Delta}=\sum_{i=-2}^{2} \frac{c_{i} X_{l}^{i}\left(k\right)}{12\Delta}.
\end{equation}
Here $X_{l}^{i}\left(k\right)$ denotes the transfer function
evaluated under the transformation $\Omega_b \to
\Omega_b\left(1+\Delta\right)$, $\Omega_c\to
\Omega_c-\Delta \Omega_b$. We find that the choice
$\Delta=0.02$ works well to evaluate the first
derivatives. We run convergence tests by doubling and halving
$\Delta$ and find that $ dX_{l}\left(k\right)/d\Delta$
has converged to $\sim 5\%$, which is more than sufficient for
our purposes. We use values 
$c_{0}=0$, $c_{\pm 1}=\pm 8$, and $c_{\pm 2}=\mp 1$ \cite{henrici}.

For the second derivatives, we use the $7$-point numerical
approximation 
\begin{equation}
     \frac{d^2 X_{l}\left(k\right)}{d\Delta^{2}}=\sum_{i=-3}^{3}
     \frac{c_{i} X_{l}^{i}\left(k\right)}{180\Delta^{2}}.
\end{equation} 
In this
case, we find that the choice $\Delta=0.066$ lies comfortably in
the zone of convergence. The corresponding coefficients are
$c_{0}=-490$, $c_{\pm 1}=270$, $c_{\pm 2}=-27$,  and $c_{\pm
3}=2$. We run convergence tests by doubling and halving $\Delta$
and find that $d^2 X_{l}\left(k\right)/ d\Delta^{2}$ has
converged to $\sim 5\%$, which is accurate enough for the work
presented here. The resulting derivative power spectra, defined
by Eqs.~(\ref{derpspec_computenot})-(\ref{derpspec_compute}), are shown in
Figs. 9--\ref{derivpspectwo}. All derivative power spectra are computed using \textsc{camb}, with ${\tau}=0$. These are then multiplied by a homogeneous reionization damping envelope with mean optical depth ${\tau}\simeq 0.086$, given by expressions in Ref. \cite{huwhite_reion}. This was done to isolate the effects of patchy decoupling, screened by a homogeneous optical depth at reionization (with $z_{\rm reion}\simeq 10.5$), from additional (smaller) anisotropies induced at reionization.

\begin{figure*}[htbp]
\includegraphics[width=6.50in]{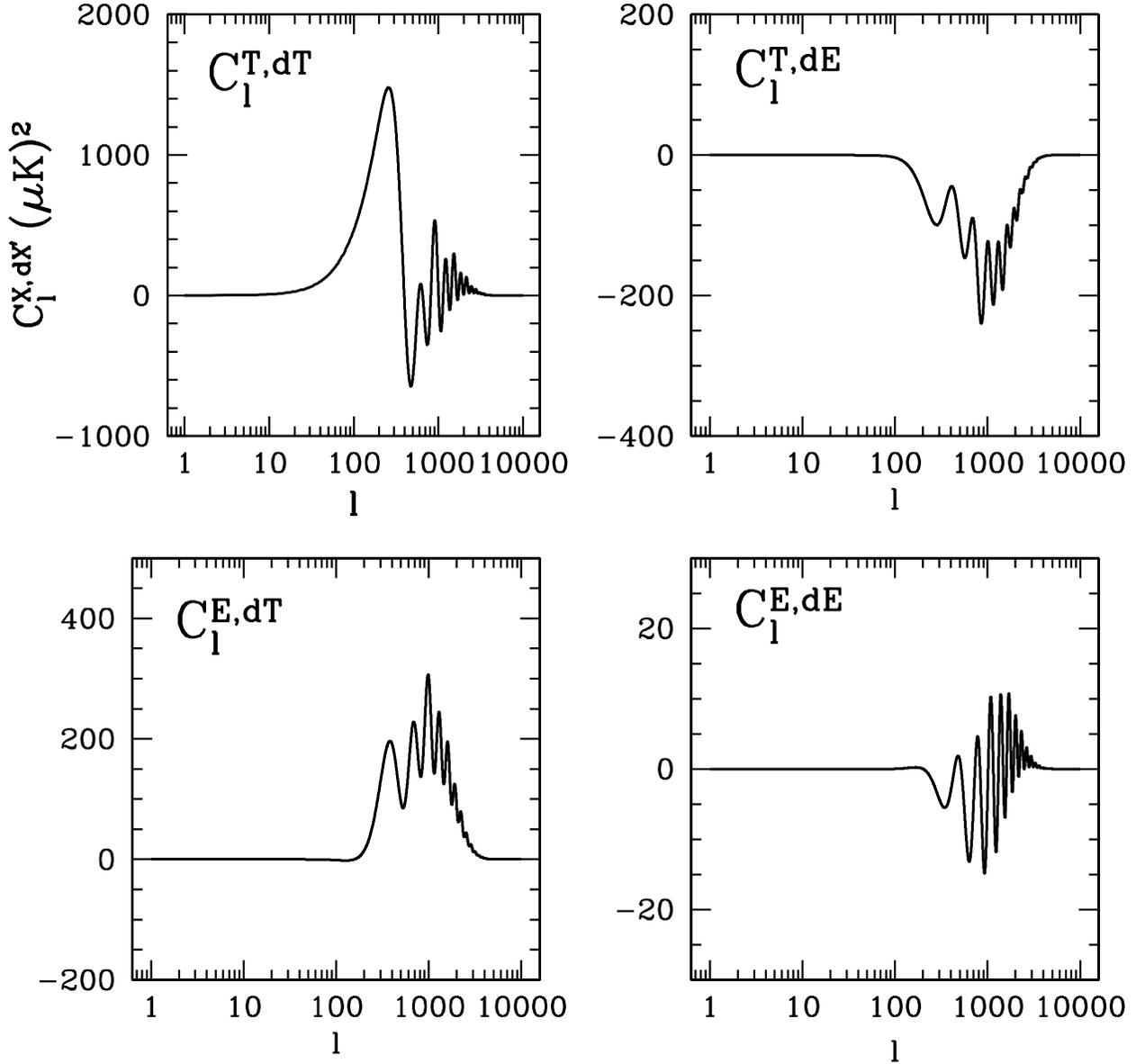}
\label{derivest}
\caption{Lowest-order first-derivative power spectra, as defined
     in Secs.~\ref{tempsec} and \ref{polaniso}. These are
     necessary to reconstruct $\Delta_{LM}$, as described in
     Sec.~\ref{qestsecbig}. We use the numerical methods of
     Appendix \ref{deriv_pspec_sec} to obtain these curves using
     a modified version of the \textsc{camb} \cite{camb} code.}
\end{figure*}

\begin{figure*}[htbp]
\includegraphics[width=6.50in]{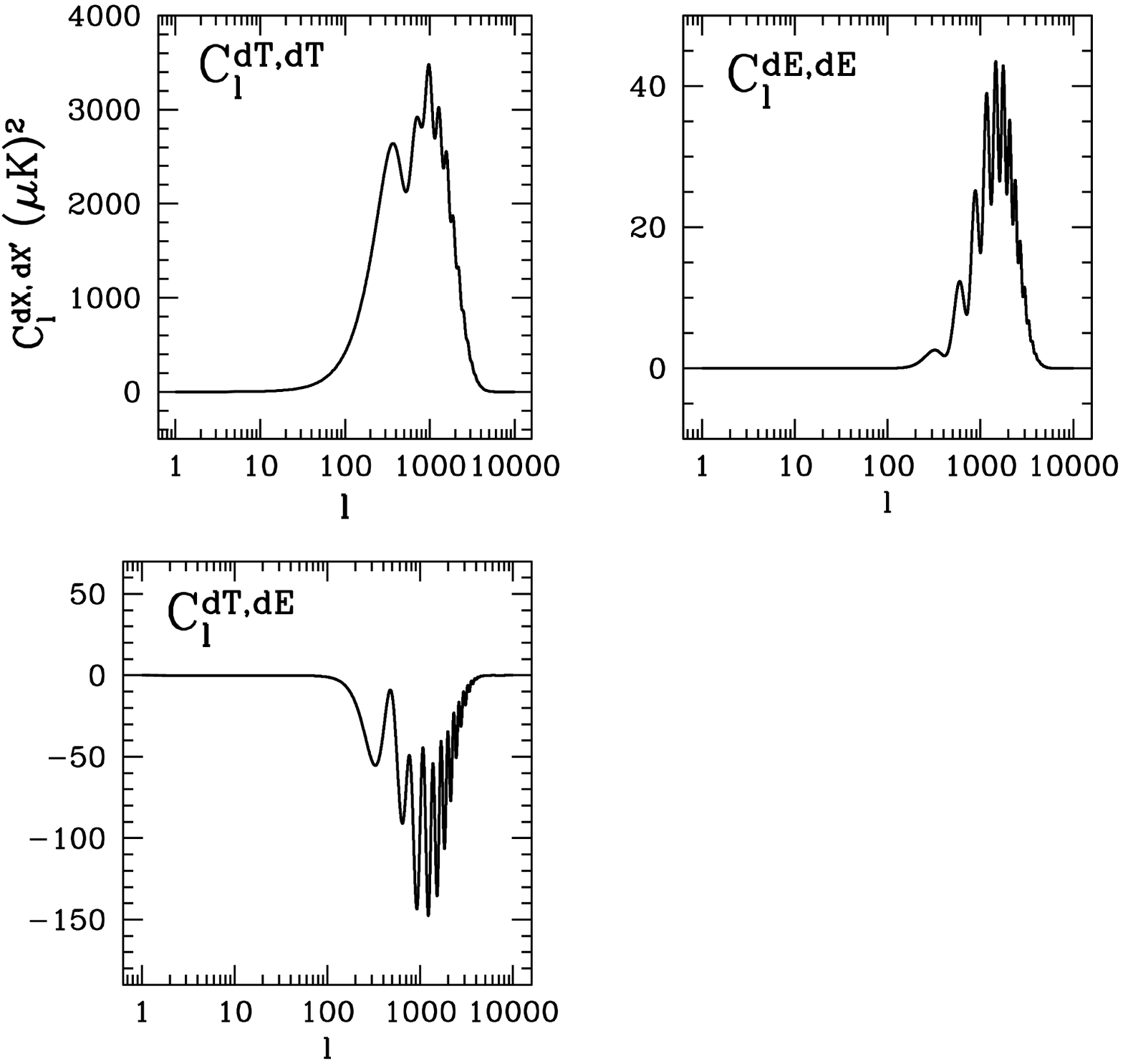}
\label{derivpspec}
\caption{Second-order first-derivative power spectra, as defined
     in Secs.~\ref{tempsec} and \ref{polaniso}. These are
     necessary to estimate the corrected power spectra
     $C_{l}^{\rm X X',(2)}$. We use the numerical
     methods of Appendix \ref{deriv_pspec_sec} to obtain these
     curves using a modified version of the \textsc{camb}
     \cite{camb} code.}
\end{figure*}

\begin{figure*}[htbp]
\includegraphics[width=6.50in]{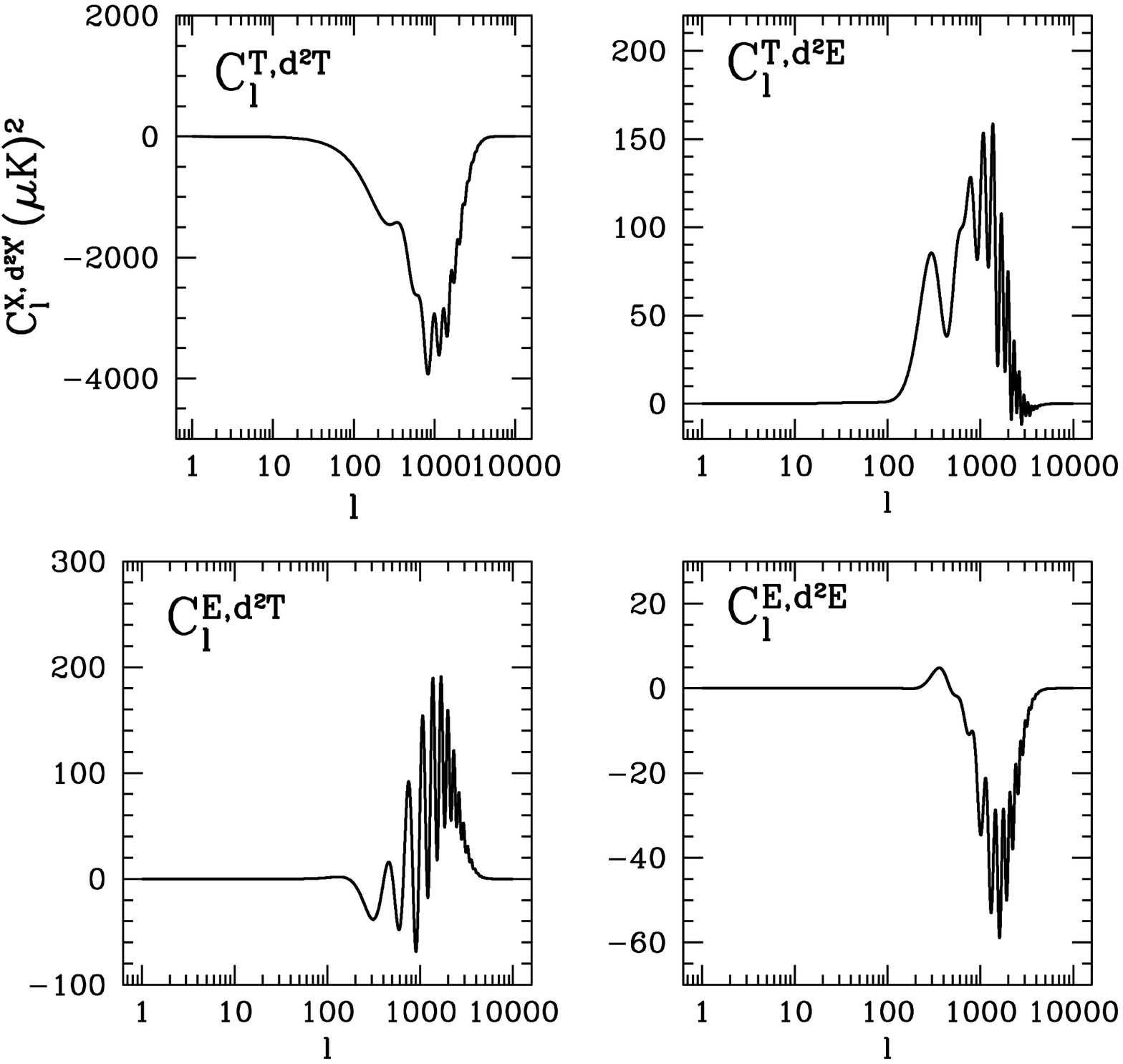}
\caption{Second-derivative power spectra, as defined in
     Secs.~\ref{tempsec} and \ref{polaniso}. These are necessary
     to estimate the corrected power spectra $C_{l}^{\rm X
     X',(2)}$. We use the numerical methods of Appendix
     \ref{deriv_pspec_sec} to obtain these curves using a
     modified version of the \textsc{camb} \cite{camb} code.} 
\label{derivpspectwo}
\end{figure*}

\section{Harmonic expansion of CMB transfer functions}
 \renewcommand{\theequation}{C\arabic{equation}}
  \setcounter{equation}{0} 
  \label{secdercorr}

The most convenient way to generalize Eq.~(\ref{tlosa}) to
include terms $\propto \Delta^{2}(\hat{n})$ is to derive
second-order corrections to
\begin{eqnarray}
     f_{LM} \equiv \int d\hat{n}\, Y_{LM}^{*}(\hat{n})
     f(\eta,\hat{n}). 
\label{flm}
\end{eqnarray}
Using the Taylor expansion in real space defined by
Eq.~(\ref{delta_exp}) and Eq.~(\ref{flm}), we obtain
\begin{eqnarray}
     f_{LM}&=&f_{LM}^{\left(1\right)}+f_{LM}^{\left(2\right)}+...,\nonumber\\
     f_{LM}^{\left(1\right)}&\equiv&\Delta_{LM}
     \frac{df^{\left(0\right)}}{d\Delta},\nonumber\\
     f_{LM}^{\left(2\right)} & \equiv & \frac{1}{2}
     \frac{d^{2}f^{\left(0\right)}} {d\Delta^{2}}
     \sum_{L'M',L^{\prime
     \prime}M''}
     \beta^{L''M''}_{L'M',LM}
     \Delta_{L'M'}
     \Delta^{*}_{L''M''}, \nonumber\\
\label{harmonic}
\end{eqnarray}
where
\begin{equation}
     \beta^{L''M''}_{L'M',LM}
      \equiv
     \xi^{L''M''}_{L'M',LM}
     K^{L}_{L',L'''}.
\end{equation}


\end{document}